\documentclass[12pt,a4paper]{amsart}

\usepackage{amsfonts,amsmath,amsthm}
\usepackage{latexsym}

\newtheorem{theorem}{Theorem}[section]
\newtheorem{corollary}[theorem]{Corollary}
\newtheorem{lemma}[theorem]{Lemma}
\newtheorem{remark}[theorem]{Remark}
\newtheorem{problem}[theorem]{Problem}
\newtheorem{Numeric}[theorem]{Numerical Evidence}

\newtheorem{definition}{Definition}[section]
\theoremstyle{definition}

\newcommand{\dst}{\displaystyle}

\newcommand \re {\mathrm{Re}\:}

\newcommand{\TT}{\ensuremath{\mathbb{T}}}
\newcommand{\R}{\ensuremath{\mathbb{R}}}

\newcommand{\Co}{\ensuremath{\mathbb{C}}}

\def \ffi {\varphi}

\def \C {\mathbb{C}}

\newcommand{\ac}{\ensuremath{\mathcal{A}}}
\newcommand{\bc}{\ensuremath{\mathcal{B}}}

\newcommand{\eb}{\ensuremath{\mathbf{e}}}
\newcommand{\fb}{\ensuremath{\mathbf{f}}}
\newcommand{\ub}{\ensuremath{\mathbf{u}}}

\def \< {\langle}
\def \> {\rangle}

\newcommand{\norm}[1]{{\left\|{#1}\right\|}}
\newcommand{\ent}[1]{{\left[{#1}\right]}}
\newcommand{\abs}[1]{{\left|{#1}\right|}}
\newcommand{\scal}[1]{{\left\langle{#1}\right\rangle}}

\begin{document}

\title[A generalized Pauli problem]{A generalized Pauli problem and an infinite family of MUB-triplets in dimension
6}

\author[P. Jaming]{Philippe Jaming}
\address{P.J.: Universit\'e d'Orl\'eans\\
Facult\'e des Sciences\\
MAPMO - F\'ed\'eration Denis Poisson\\ BP 6759\\ F 45067 Orl\'eans Cedex 2\\
France} \email{philippe.jaming@univ-orleans.fr}

\author[M. Matolcsi]{M\'at\'e Matolcsi}
\address{M. M. and M.W.: Alfr\'ed R\'enyi Institute of Mathematics,
Hungarian Academy of Sciences POB 127 H-1364 Budapest, Hungary
Tel: (+361) 483-8302, Fax: (+361) 483-8333}
\email{matomate@renyi.hu, mweiner@renyi.hu}

\thanks{M. Matolcsi was supported by OTKA Grants No. PF-64061, T-04930, the
Junior Fellowship of ESI (Wien, February 2008), and a Visiting
Professorship of the University of Orleans, October 2007.}

\author[P. M\'ora]{P\'eter M\'ora}
\address{P. M. (and M. M. part time): BME Department of Analysis,
Egry J. u. 1,H-1111 Budapest, Hungary} \email{morapeter@gmail.com}

\author[F. Sz\"oll\H osi]{Ferenc Sz\"oll\H osi}
\address{F. Sz.: Central European University (CEU),
Institute of Mathematics and its Applications, H-1051, N\'ador u.
9., Budapest, Hungary}\email{szoferi@gmail.com}

\author[M. Weiner]{Mih\'aly Weiner}


\begin{abstract}
We exhibit an infinite family of {\it triplets} of mutually
unbiased bases (MUBs) in dimension 6. These triplets involve the
Fourier family of Hadamard matrices, $F(a,b)$. However, in the
main result of the paper we also prove that for any values of the
parameters $(a,b)$, the standard basis and $F(a,b)$ {\it cannot be
extended to a MUB-quartet}. The main novelty lies in the {\it
method} of proof which may successfully be applied in the future
to prove that the maximal number of MUBs in dimension 6 is three.
\end{abstract}

\maketitle



\section{Introduction}

The notion of mutually unbiased bases (MUBs) emerged in the
literature of quantum mechanics in 1960 in the works of Schwinger
\cite{Sc}. It now constitutes a basic concept of Quantum
Information Theory and plays an essential role in
quantum-tomography \cite{Iv,WF}, quantum criptography
\cite{BPT,BB,Re}, the mean king problem \cite{AE} as well as in
constructions of teleportation and dense coding schemes
\cite{wer}.

Recall that two orthonormal bases of $\Co^d$,
$\ac=\{\eb_1,\ldots,\eb_d\}$ and $\bc=\{\fb_1,\ldots,\fb_d\}$ are
said to be \emph{unbiased} if, for every $1\leq j,k\leq d$,
$\abs{\scal{\eb_j,\fb_k}}=\dst\frac{1}{\sqrt{d}}$. A set
$\bc_0,\ldots\bc_m$ of orthonormal bases is said to be
\emph{mutually unbiased} if any two of them are unbiased. It is
well-known (see e.g. \cite{BBRV,BBELTZ,DGS,Ho,KL,WF}) that the
number of mutually unbiased bases in $\Co^d$ cannot exceed $d+1$.
It is also known that $d+1$ such bases can be constructed if the
dimension $d$ is a prime or a prime power (see e.g.
\cite{BBRV,Com0,Com1,Com2,Iv,KR,WF}). Apart from this, very little
is known except for the fact that there are always $p+1$ mutually
unbiased bases in $\C^d$ where $p$ is the smallest prime divisor
of $d$. Thus, the first case where the largest number of mutually
unbiased bases is unknown is $d=6$:

\begin{problem}\label{MUB6problem}\ \\
What is the maximal number of pairwise mutually unbiased bases in
$\Co^6$?
\end{problem}

Although this famous open problem has received considerable
attention over the past few years
(\cite{BBELTZ,config,ujbrit,Msz,Skinner}), it remains wide open.
Since $6=2\times 3$, we know that there are at least $3$ mutually
unbiased bases in $\C^6$, but so far tentative numerical evidence
\cite{config,ujbrit,numerical,Za} suggests that there are no more
than 3, a fact apparently first conjectured by Zauner \cite{Za}.

One reason for the slow progress is that mutually unbiased bases
are naturally related to \emph{complex Hadamard matrices} (and the
classification of such matrices in dimension 6 seems to be very
difficult). Indeed, if the bases $\bc_0,\ldots,\bc_m$ are mutually
unbiased we may identify each
$\bc_l=\{\eb_1^{(l)},\ldots,\eb_d^{(l)}\}$ with the \emph{unitary}
matrix $U_j=\ent{\scal{\eb_k^{(l)},\eb_j^{(1)}}}_{1\leq j,k\leq
d}$, {\it i.e.} the $k$-th column consists of the coordinates of
the $k$-th vector of $\bc_j$ in the bases $\bc_0$. (Throughout the
paper the scalar product $\scal{.,.}$ of $\C^d$ is linear in the
first variable and conjugate-linear in the second). With this
convention, $U_1=Id$ the identity matrix and all other matrices
are unitary and have entries of modulus $1/\sqrt{d}$. Such
matrices are called \emph{complex Hadamard matrices}. It is clear
that the existence of a family of mutually unbiased bases
$\bc_0,\ldots,\bc_m$ is thus equivalent to the existence of a
family of complex Hadamard matrices $H_1,\ldots,H_m$ such that for
all
 $1\leq j\not=k\leq
m$, $H_j^*H_k$ is again a complex Hadamard matrix. In such a case
we will say that these complex Hadamard matrices are {\it mutually
unbiased}.

A complete classification of complex Hadamard matrices is only
available up to dimension 5 (see \cite{haagerup}). The
classification in dimension 6 is still out of reach despite recent
efforts \cite{BN,Msz,Skinner}. This is one of the reasons for
Problem \ref{MUB6problem} to be difficult.

A natural question that arises in this context is that given two
unbiased orthonormal bases, does there always exist a third
orthonormal basis that is unbiased to the first two? Or,
equivalently, given a complex Hadamard matrix $H$, does there
always exist another one $G$ that is unbiased to $H$? The answer
is negative in such generality. It was recently proved in
\cite{ujbrit} that for the matrix $S_6$ (cf. \cite{karol} for the
notation) there exists no complex Hadamard matrix unbiased to it.
A less restrictive question is the following:

\begin{problem}\label{pb:mubextend}\ \\
Given a Hadamard matrix $H$, does there always exist {\em some}
unbiased vectors to $H$?
\end{problem}

At this stage, it may be worth recalling some invariants of this
problem. Assume $H_1,\ldots,H_m$ are mutually unbiased $n\times n$
Hadamard matrices, and let $D,D_1,\ldots,D_m$ be unitary $n\times
n$ diagonal matrices $P,P_1,\ldots,P_m$ be $n\times n$ permutation
matrices. Then $DPH_iP_iD_i$ are still Hadamard matrices and are
still mutually unbiased. We will say that $DPH_i^{(*)}P_iD_i$ is
\emph{equivalent} to $H_i$ where the superscript $(*)$ is a choice
(the same for all matrices) between complex conjugation or
nothing. Note that the effect of this operation is that we may
assume that the first \emph{row} of each Hadamard matrix is
$d^{-1/2}[1,\ldots,1]$ and that the first \emph{column of $H_1$}
is $d^{-1/2}[1,\ldots,1]$. We may further \emph{order} the
remaining rows \emph{and} columns of $H_1$.

Further, note that if an $n$-dimensional vector
\begin{equation}\label{udef}
\ub=\dst\frac{1}{\sqrt{6}}(1,e^{2i\pi\phi_1},e^{2i\pi\phi_2},e^{2i\pi\phi_3},e^{2i\pi\phi_4},e^{2i\pi\phi_5})
\end{equation}
is unbiased to the standard basis and to the $n-1$ first columns
of a Hadamard matrix, then it is automatically unbiased to the
last one. Hence, we have $n-1$ unbiased-criteria to be satisfied
for the $n-1$ parameters $\phi_j$. In generic situations we
therefore expect a finite number of solutions to arise. We know of
a non-generic example (in dimension 4) where infinitely many
unbiased vectors arise, but of no examples where the number of
such vectors is {\it zero}.

\medskip

Moving back to the $6$-dimensional case we note the significance
of Problem \ref{pb:mubextend}. If for a certain complex Hadamard
matrix $H$ the number of unbiased vectors is less than 30, then
the MUB-pair $\{Id,H\}$ can obviously not be extended to a full
set of 7-MUBs (because we would need at least 30 vectors to form
another 5 bases).

Further, it is easy to see that a vector $\ub$ of the form
\eqref{udef} is unbiased to the columns of $Id$ and of $H$ if and
only if the mapping
\begin{equation}\label{nummax}
\mathcal{H}(\phi_1,\phi_2,\phi_3,\phi_4,\phi_5)=\sum_{j=1}^6
|\langle {\bf u , h_j} \rangle |,
\end{equation}
(where $\bf{h_j}$ denote the columns of $H$) has a global maximum
at the point $(\phi_1,\phi_2,\phi_3,\phi_4,\phi_5)$ in $\TT^5$.

Therefore, a natural way to search numerically for unbiased
vectors $\ub$ is to start from a random point of the parameter
space $(\phi_1,\phi_2,\phi_3,\phi_4,\phi_5)$ in $\TT^5$ and find
local maxima of $\mathcal{H}$ defined in \eqref{nummax}. It is
plausible to expect that if we run our numerical search many times
we will find most, or indeed all, unbiased vectors $\ub$ in this
manner (as well as finding possible other local maxima which we
simply discard). There is no guarantee, of course, and we will
need to back up our numerical evidence with rigorous mathematical
statements.

\medskip

In most of this paper, we will focus on $H$ belonging to the
``Fourier family''. Let us recall (cf. \cite{karol}) that this is
the two-parameter family of complex Hadamard matrices $F(a,b)$
defined by
\begin{equation}\label{fab}
F(a,b)=\frac{1}{\sqrt{6}}\left[\begin{array}{rrrrrr}
 1 & 1 & 1 & 1 & 1 & 1 \\
 1 & -\omega^2 x & \omega  & -x & \omega^2  & -\omega x\\
 1 & \omega y& \omega^2 & y & \omega & \omega^2 y\\
 1 & -1 & 1 & -1 & 1 & -1 \\
 1 & \omega^2 x& \omega& x & \omega^2 & \omega x\\
 1 & -\omega y & \omega^2  & -y & \omega  & -\omega^2 y
\end{array}\right],
\end{equation}
where $x=e^{2i\pi a}$, $y=e^{2i\pi b}$ and $\omega=e^{2i\pi/3}$.

Note that $F(0,0)=\mathcal{F}_6$ is the standard Fourier matrix.
We may thus see the task of finding unbiased vectors to the
standard and the $F(a,b)$ bases as a perturbation of the so-called
Pauli Problem. Recall that Pauli asked whether a function $f$ in
$L^2(\R^d)$ is uniquely determined by its modulus $|f|$ and the
modulus of its Fourier Transform $|\widehat{f}|$ ({\it see e.g.}
\cite{Ja} or \cite{Cor} for some results and further references).
The discrete analogue of this problem, {\it i.e.} the problem of
finding finite sequences of complex numbers of modulus one $(a_j)$
such that their Fourier transforms have also modulus one (such
sequences are called \emph{biunimodular sequences}) has been
considered {\it e.g.} in \cite{BaBj,BP}. Our problem can thus be
seen as a perturbation of the discrete case in dimension $6$.

For the particular case of $H=\mathcal{F}_6$, a full analytical
solution of Problem \ref{pb:mubextend} is actually known
\cite{BaBj,grassl}: there are exactly 48 vectors, normalized as in
\eqref{udef}, that are unbiased with respect to
$\{Id,\mathcal{F}_6\}$ and one can form 16 different orthonormal
bases $C_1, \dots C_{16}$ out of them. However, no pair of bases
$(C_i, C_j)$ are unbiased with respect to each other, which means
that no triplet $\{I, \mathcal{F}_6,C\}$ can be extended to a
mutually unbiased-quartet $\{I, \mathcal{F}_6,C,D\}$ (see
\cite{grassl,ujbrit}).  What happens if we set $H=F_6(a,b)$ for
some generic values $a,b$? We heuristically expected that in such
a case significantly less than 48 unbiased vectors $\bf{u}$ should
arise. We also expected that only in exceptional cases should
there exist a basis $C$ built from these unbiased vectors. These
heuristics turned out to be false.\footnote{The authors are
grateful for W.\ Bruzda for making the first computer search, with
a method other than maximizing expression \eqref{nummax}, which
indicated that the number of unbiased vectors is more than $40$ on
average.}

We ran the numeric search of finding local maxima of expression
\eqref{nummax} for several values of $a,b$. The results were both
surprising and overwhelmingly convincing.

\begin{Numeric}\label{numevi}\ \\
For {\em any values} of the parameters $(a, b)$, the number of
vectors unbiased to the identity matrix and $F(a,b)$ {\em is 48}.
For {\em generic} values $(a, b)$ there are {\em 8} different {\em
orthonormal bases} $C_1(a,b), \dots C_8(a,b)$ that can be formed
out of these 48 vectors. For some exceptional values of $(a,b)$
there are more such bases: for $(a,b)=(0, 0)$ there are 16, while
for $(a,b)=(1/6,0)$ there are 70 such bases\footnote{The exact
number 70 is given in \cite{ujbrit}}. However, {\em no mutually
unbiased triplet} of the form $(Id, F(a,b), C)$ {\em can be
extended} by a further basis to form a mutually unbiased quartet
$(Id, F(a,b), C, D)$.
\end{Numeric}

We will back up most of these numerical data by {\it rigorous
analytic results} in subsequent sections.\footnote{The authors of
\cite{ujbrit} have used the technique of Gr\"obner bases to prove
that the number of unbiased vectors is indeed 48 for several (but
{\it finitely many}) tested values of $(a,b)$. In fact, in
\cite{ujbrit} several members of {\em all} known Hadamard families
are tested, not only the Fourier family. However, the techniques
of this paper have the advantage that they enable us to reach
rigorous conclusions about the {\it whole family} $F(a,b)$ and not
just the tested values of the parameters.}

\medskip

In Section \ref{sec:infinite} we construct an {\it infinite family
of MUB-triplets in analytic form} involving the Fourier family of
Hadamard matrices $F(a,b)$, with $a=0$ (and some restrictions on
$b$). We have recently been informed by G. Zauner that his work
\cite{Za} also includes an infinite family of MUB-triplets,
although the formulas are not made explicit. As the beautiful
construction of \cite{Za} is scarcely known and it is originally
written in German we decided to provide an English version of it
in the Appendix of \cite{arxiv}. We will show, however, that
Zauner's family is not equivalent to ours.

\medskip

One may think that the emergence of an infinite family of
MUB-triplets is a major step towards finding a MUB-quartet in
dimension 6. On the contrary, we prove the following:

\begin{theorem}\label{no4s}\ \\
None of the pairs $\bigl(Id, F(a,b)\bigr)$ of mutually unbiased
orthonormal bases can be extended to a quartet $\bigl(Id,
F(a,b),C,D\bigr)$ of mutually unbiased orthonormal bases.
\end{theorem}

While this can be disappointing for some, we believe that this is
a {\it breakthrough result} of the paper in that the method we
apply here may later be generalized to {\it settle Problem
\ref{MUB6problem}} and prove that the maximal number of mutually
unbiased orthonormal bases in dimension 6 is three.

\medskip

The paper is organized as follows: Section \ref{sec:infinite} is
devoted to characterizing vectors unbiased to the standard basis
and $F(a,b)$, and the construction of an infinite one-parameter
family of MUB triplets. In Section \ref{sec:no4mubs} we prove
Theorem \ref{no4s}. Finally, in Section \ref{sec:theory} we
attempt to offer some general theoretical reasons behind our
results, other than just the sheer numbers and formulae.

\section{An infinite family of MUB-triplets involving
$F(a,b)$}\label{sec:infinite}

In this section we first describe a reduced system of equations
for any vector $\bf u$ unbiased to the bases $A=Id$ and
$B=F(a,b)$. However, we can only obtain some particular solutions
in closed analytic form in the special case $a=0$. Nevertheless,
these explicit formulae give a {\it rigorous proof of the
existence of an infinite 1-parameter family of MUB-triplets}
involving $A=Id$ and $B=F(0,b)$. Recall, that the numerical
evidence actually suggests existence of MUB-triplets for {\it all}
$a,b$, with $A=Id$ and $B=F(a,b)$, i.e. a {\it two-parameter
family}. We cannot give a rigorous argument in such generality.

\subsection{A reduced system of equations for unbiased vectors}\

We begin with a useful lemma about vectors unbiased with the
Fourier basis in $\C^3$:

\begin{lemma}\label{lem:1}\ \\
Let $\omega=e^{2i\pi/3}$ and let $\alpha,\beta,\gamma\in\C$. Then
\begin{equation}
\label{eq:lem1} \left\{\begin{matrix}
|\alpha&+&\beta&+&\gamma|^2&=&6\\
|\alpha&+&\beta\omega&+&\gamma\omega^2|^2&=&6\\
|\alpha&+&\beta\omega^2&+&\gamma\omega|^2&=&6
\end{matrix}\right.
\end{equation}
if and only if
$$
\left\{\begin{matrix}
|\alpha|^2+|\beta|^2+|\gamma|^2&=&6\\
\alpha\bar\beta+\beta\bar\gamma+\gamma\bar\alpha&=&0
\end{matrix}\right..
$$
\end{lemma}

\begin{proof}
One easily sees that \eqref{eq:lem1} is equivalent to
$$
\left\{\begin{matrix} |\alpha|^2+|\beta|^2+|\gamma|^2
&+&2\mbox{Re}\bigl(\alpha\bar\beta+\beta\bar\gamma+\gamma\bar\alpha\bigr)&=&6\\
|\alpha|^2+|\beta|^2+|\gamma|^2
&+&2\mbox{Re}\bigl(\omega^2\alpha\bar\beta+\omega^2\beta\bar\gamma+\omega^2\gamma\bar\alpha\bigr)&=&6\\
|\alpha|^2+|\beta|^2+|\gamma|^2
&+&2\mbox{Re}\bigl(\omega\alpha\bar\beta+\omega\beta\bar\gamma+\omega\gamma\bar\alpha\bigr)&=&6
\end{matrix}\right..
$$
Then, using the fact that $1+\omega+\omega^2=0$ and adding all
three equations, we see that this is equivalent to
$$
\left\{\begin{matrix}
|\alpha|^2+|\beta|^2+|\gamma|^2&=&6\\
\mbox{Re}\bigl(\alpha\bar\beta+\beta\bar\gamma+\gamma\bar\alpha\bigr)&=&0\\
\mbox{Re}\bigl(\omega\alpha\bar\beta+\omega\beta\bar\gamma+\omega\gamma\bar\alpha\bigr)&=&0\\
\end{matrix}\right..
$$
We conclude by noticing that $\mbox{Re}(z)=0$ and
$\mbox{Re}(\omega z)=0$ if and only if $z=0$.
\end{proof}

Let us now assume that ${\bf{u}}\in\C^6$ is a unit-norm vector
that is unbiased to the standard basis:
$$
{\bf{u}}=\frac{1}{\sqrt{6}}(1,\overline{c}_1,\overline{c}_2,\overline{c}_3,\overline{c}_4,\overline{c}_5),\qquad
|c_1|=|c_2|=|c_3|=|c_4|=|c_5|=1,
$$ where the conjugate signs are introduced for later convenience of calculations.

Recalling the notation $x=e^{2i\pi a}$, $y=e^{2i\pi b}$, the
vector $\bf{u}$ is further unbiased with respect to the
generalized Fourier basis $F(a,b)$, if and only if
\begin{eqnarray}
|1+c_1+c_2+c_3+c_4+c_5|&=&{\sqrt{6}}\label{eq:a}\\
|1-\omega^2 xc_1+\omega yc_2-c_3+\omega^2 xc_4-\omega yc_5|&=&{\sqrt{6}}\label{eq:b}\\
|1+\omega c_1+\omega^2 c_2+c_3+\omega c_4+\omega^2 c_5|&=&{\sqrt{6}}\label{eq:c}\\
|1-xc_1+yc_2-c_3+xc_4-yc_5|&=&{\sqrt{6}}\label{eq:d}\\
|1+\omega^2 c_1+\omega c_2+c_3+\omega^2 c_4+\omega c_5|&=&{\sqrt{6}}\label{eq:e}\\
|1-\omega xc_1+\omega^2 yc_2-c_3+\omega xc_4-\omega^2
yc_5|&=&{\sqrt{6}}\label{eq:f}.
\end{eqnarray}

\vskip 0.3 truecm

Applying Lemma \ref{lem:1} to Equations
\eqref{eq:a},\eqref{eq:c},\eqref{eq:e}, we obtain
\begin{eqnarray}
|1+c_3|^2+|c_1+c_4|^2+|c_2+c_5|^2&=&6\label{eq:1a}\\
(1+c_3)\overline{(c_1+c_4)}+(c_1+c_4)\overline{(c_2+c_5)}&&\notag\\
+(c_2+c_5)\overline{(1+c_3)}&=&0\label{eq:1c}
\end{eqnarray}
while applying it to Equations
\eqref{eq:b},\eqref{eq:d},\eqref{eq:f} we obtain the following:
\begin{eqnarray}
|1-c_3|^2+|x(-c_1+c_4)|^2+|y(c_2-c_5)|^2&=&6\label{eq:1b}\\
(1-c_3)\overline{x(c_4-c_1)}+x(c_4-c_1)\overline{y(c_2-c_5)}&&\notag\\
+y(c_2-c_5)\overline{(1-c_3)}&=&0.\label{eq:1d}
\end{eqnarray}
But, using the fact that $c_1,\ldots,c_5$ are all of modulus $1$,
we see that Equation \eqref{eq:1a} is equivalent to
\begin{equation}
\label{eq:1e}
\mbox{Re}(c_3+c_1\overline{c_4}+c_5\overline{c_2})=0.
\end{equation}
Similarly, as $|x|=|y|=1$, \eqref{eq:1b} reads
$|1-c_3|^2+|c_4-c_1|^2+|c_2-c_5|^2=6$ which also reduces to
\eqref{eq:1e}.

We have thus proved the following lemma:

\medskip

\begin{lemma}\label{lem:ph1117}\ \\
A vector
${\bf{u}}=\frac{1}{\sqrt{6}}(1,\overline{c}_1,\overline{c}_2,\overline{c}_3,\overline{c}_4,\overline{c}_5))$
is unbiased to both the standard basis and the generalized Fourier
basis $F(a,b)$ if and only if all $c_j$ have absolute value 1, and
the following conditions are fulfilled
$$
\left\{\begin{matrix}
\re (c_3+c_1\overline{c_4}+c_5\overline{c_2})=0&\qquad\eqref{eq:1e}\\
(1+c_3)\overline{(c_1+c_4)}+(c_1+c_4)\overline{(c_2+c_5)}&&\notag\\
+(c_2+c_5)\overline{(1+c_3)}=0&\qquad\eqref{eq:1c}\\
(1-c_3)\overline{x(c_4-c_1)}+x(c_4-c_1)\overline{y(c_2-c_5)}&&\notag\\
+y(c_2-c_5)\overline{(1-c_3)}=0&\qquad\eqref{eq:1d}
\end{matrix}\right..
$$
\end{lemma}

\medskip

\begin{remark}\rm
The most natural way to prove all statements indicated by
Numerical Evidence \ref{numevi} would be to {\it solve this system
of equations} and show that there are exactly 48 solution vectors
for any choice of $a,b$. This would provide an {\it exhaustive
list of MUB-triplets} involving the generalized Fourier matrices
$F(a,b)$. Unfortunately, we have been unable to fulfill this task
in this generality and we are not able to give {\it all} solutions
in closed analytic form.
\end{remark}

\subsection{An infinite family of MUB-triplets involving $F(0,b)$}
In order to obtain analytic formulae for \emph{some} of the
arising MUB-triplets we need to restrict our attention to the case
$a=0$. Even in this case the calculations are rather long and
cumbersome, and not very instructive. The full details are
presented in the Appendix of \cite{arxiv}, and we only include the
final result here. Note that all emerging formulae are {\it
explicit} so that the correctness of the result can be checked
(most conveniently by computer algebra) without going through the
detailed calculations.

\begin{theorem}\label{thmmubtriplet}
Assume that $\frac{1}{2}\arcsin\frac{\sqrt{5}}{3}\leq
t\leq\frac{\pi}{2}-\frac{1}{2}\arcsin\frac{\sqrt{5}}{3}$.
Introduce the following variables:
\begin{equation}
\label{psichic3} \psi=\arccos\frac{\sqrt{2+\cos 2t}}{2}, \ \ \ \ \
c_3=-\frac{\cos 2t}{2}+ i\left(1-\frac{\cos^22t}{4}\right)^{1/2}
\end{equation}
\begin{equation}
\beta = \arccos \frac{\sqrt{3+\sin^22t}+3\sqrt{9\sin^22t-5}}{8\sin
2t}.\label{beta}
\end{equation}
Then define $\ffi$ and $\tilde{\varphi}$ by the equations
\begin{equation}
\label{cffi} \cos\ffi=\frac{-\cos^2\psi\cos
t+\cos(\beta+\psi)\sin\psi\sin t} {\sin\beta\sin 2t},
\end{equation}
\begin{equation}
\label{sffi} \sin\ffi=\frac{\sin\psi\cos\psi\cos
t-\sin(\beta+\psi)\sin\psi\sin t} {\sin\beta\sin 2t},
\end{equation}
and
\begin{eqnarray}
\cos{\tilde\varphi} &=&\frac{-\sin t\sin^2\psi+\cos\psi\cos
t\sin(\beta+\psi)}{\sin\beta\sin 2t}
\label{cffiuj}\\
\sin{\tilde\varphi} &=&\frac{\cos \psi\cos(\beta+\psi)\cos
t-\cos\psi\sin t\sin\psi}{\sin\beta\sin 2t}. \label{sffiuj}
\end{eqnarray}
 Write $\eta=e^{it}$, $\nu=e^{i\ffi}$, $\xi=e^{i\tilde{\varphi}}$, and $b=\frac{1}{2\pi}\beta$. Finally
define $C(t)$ to be the orthonormal basis given by columns of the
matrix
$$
\frac{1}{\sqrt6} \left[
\begin{array}{cccccc}
 1 & 1 & 1 & 1 & 1 & 1 \\
 \bar{c}_3 &  \bar{c}_3\omega ^2 &   \bar{c}_3\omega & -\bar{c}_3 & -
\bar{c}_3\omega ^2 & -  \bar{c}_3\omega \\
 \bar{\nu }\eta   &   \bar{\nu }\eta  \omega &  \bar{\nu }\eta
\omega ^2 & i   \bar{\xi }\eta & i   \bar{\xi }\eta  \omega & i
\bar{\xi }\eta  \omega ^2 \\
 \bar{c}_3 & \bar{c}_3 & \bar{c}_3 & -\bar{c}_3 & -\bar{c}_3 & -\bar{c}_3 \\
 1 & \omega ^2 & \omega  & 1 & \omega ^2 & \omega  \\
 \bar{\nu } \bar{\eta } & \bar{\nu }   \bar{\eta }\omega &
\bar{\nu } \bar{\eta }\omega ^2  & -i  \bar{\xi }\bar{\eta } & -i
\bar{\xi }\bar{\eta }\omega & -i   \bar{\xi }\bar{\eta }\omega ^2
\end{array}
\right]
$$
Then the standard basis, the generalized Fourier basis
$F\bigl(0,b(t)\bigr)$ and the basis given by $C(t)$ are mutually
unbiased.
\end{theorem}

\begin{remark}\rm
This theorem exhibits an infinite family of MUB triplets in terms
of a parameter $t$. Each member of the family contains the
standard basis and one member of the family $F(0,b(t))$. However,
the dependence of $t$ on $b$ is only implicit and seems unsolvable
in closed form. Note also that $b(t)$ does not take the value $0$.

We do not claim that we have found all MUB-triplets containing the
standard basis and $F\bigl(0,b(t)\bigr)$, but only one such
triplet. Actually, Numerical Evidence \ref{numevi} shows that
there exist other solutions, but we have been unable to describe them all
analitically.

Note also, that the family of MUB-triplets above is different from
the one presented in \cite{Za}. This fact is shown in the Appendix
of \cite{arxiv}.
\end{remark}

\begin{remark}\rm
It is natural to ask whether $C(t)$ provides a new family of
complex Hadamard matrices of order 6. This is not the case as
$C(t)$ also belongs to the the generalized Fourier family
$F(a,b)$. This can easily be seen by dephasing the first column
and properly reordering the remaining ones.
\end{remark}

\section{No quartet of mutually unbiased bases involving the identity and $F(a,b)$}\label{sec:no4mubs}

In this section we prove Theorem \ref{no4s}, i.e. the {\it
non-existence} of quartets of mutually unbiased bases of the form
$(Id,F(a,b),C,D)$ for any values of $a,b$. This will be done via a
discretization scheme and an exhaustive computer search after
establishing proper estimates of the error terms. We believe that
the method can be {\it generalized} in the future to prove that
the maximal number of MUBs in dimension 6 is three.

\begin{proof}[Proof of Theorem \ref{no4s}] Let us briefly describe the basic idea and turn to the
details later. The proof proceeds by contradiction: assume there
exists a MUB-quartet $(Id,F(a,b),C,D)$. First, as described in the
introduction, we may take advantage of the equivalence relations
of Hadamard matrices to reduce the range of parameters. A priori,
$(a,b)$ is any point in the square $[0,1)^2$. But, due the
equivalences described in \cite{BBELTZ} we can assume without loss
of generality that $(a,b)$ lies in the triangle $T$ with vertices
$(0,0), (1/6,0)$ and $(1/6,1/12)$ which is a {\it fundamental
region} (see \cite{BBELTZ} for details).

Next, $e^{2i\pi a}$, $e^{2i\pi b}$ and all entries of $\sqrt{6}C$
and $\sqrt{6}D$ are unimodular complex numbers. We will thus
approximate them by $N$-th roots of unity and replace the matrix
$F(a,b)$ by a matrix $F(\tilde a,\tilde b)$ and $\sqrt{6}C$,
$\sqrt{6}D$ by matrices $\sqrt{6}\tilde C,\sqrt{6}\tilde D$ with
entries exclusively $N$-th roots of unity. Of course, in doing so,
we will destroy the main features of $C,D$: namely, $\tilde C$ and
$\tilde D$ are neither unitary nor unbiased to $F(\tilde a,\tilde
b)$ (or to each other) anymore. However, if $N$ is large enough,
then $(Id,F(\tilde a,\tilde b),\tilde C,\tilde D)$ will still {\it
approximately} be a MUB-quartet. Moreover, the bounds in these
approximations can be precisely controlled. It turns out that if
$N$ is large enough, an {\it exhaustive} computer search shows
that no quartet of matrices satisfies the prescribed bounds. This
means that the hypothetical quartet $(Id,F(a,b),C,D)$ cannot
exist. The code of the computer algorithm and the full
documentation of the results are available at the web-page
\cite{web}. The running time of the code was about 6 hours on a
computer with a 3,2 GHz CPU.

\medskip

Let us now describe the details. Let $N$ be an integer. We
partition the interval $[0,1)$ into $N$ sub-intervals
$I_0^{(N)},I_1^{(N)},\ldots, I_{N-1}^{(N)}$ of equal length, {\it
i.e.} $I_j^{(N)}=[j/N,(j+1)/N)$, and denote by
$r_j^{(N)}=(j+1/2)/N$ the midpoint of $I_j^{(N)}$. Now, if $a$
(resp. $b$) fall in some interval $I_j^{(N)}$ (resp. $I_m^{(N)}$),
we then replace $a$ by $\tilde a=r_j^{(N)}$ (resp. $b$ by $\tilde
b=r_m^{(N)}$). When doing so we must keep in mind that the actual
value of $a$ (resp. $b$) can lie anywhere in the {\it interval} $I_j^{(N)}$ (resp. $I_m^{(N)}$).

Next, recall that there is no loss of generality in assuming that
all vectors in all appearing bases have first coordinate
$1/\sqrt{6}$. All other entries have modulus $1/\sqrt{6}$. This
has the consequence that all appearing scalar products throughout
this section have the form $\langle \mathbf{u}, \mathbf{v}\rangle
=\frac{1}{6}(1+\sum_{k=1}^5 e^{2i\pi (\phi_k-\psi_k)})$.

Denote the entries of $\sqrt{6}C, \sqrt{6}D$ as $c_{k,j}=e^{2i\pi
\gamma_{k,j}}$ and $d_{k,j}=e^{2i\pi \rho_{k,j}}$ with
$\gamma_{k,j},\rho_{k,j}\in [0,1)$, and the indexing being set as
$0\le k,j\le5$. Thus, for $k=0$ we have
$\gamma_{k,j}=\rho_{k,j}=0$, and for $k\geq 1$ each
$\gamma_{k,j},\rho_{k,j}$ falls into some interval
$I_\ell^{(N')}$, where $N'$ is another integer (for clarity of
notation the dependence of $\ell$ on $\gamma_{k,j},\rho_{k,j}$ has
been dropped). We define $\sqrt{6}\tilde C, \sqrt{6}\tilde D$ by
replacing these entries by $r_\ell^{(N')}$. It turns out that we
can take $N'$ smaller than $N$, which saves a lot of computing
time. Actually, our search was carried out with $N=180$ and
$N'=19$.

Finally, the algorithm runs in two steps. In the first one, we
seek all vectors $\tilde{\mathbf{u}}$ of the form
\begin{equation}
\label{eq:approx}
\tilde{\mathbf{u}}=\frac{1}{\sqrt{6}}(1,e^{2i\pi(j_1+1/2)/N'},\ldots
,e^{2i\pi(j_5+1/2)/N'})
\end{equation}
that are ``almost" unbiased to $F(\tilde a,\tilde b)$. These
vectors are the candidates for the columns of $\tilde{C},
\tilde{D}$. The second step then consists of constructing ``almost"
orthonormal bases $\tilde{C}, \tilde{D}$ out of these vectors, and
checking whether those can possibly be ``almost" unbiased to each
other. Of course, all the ``almost" terms above need to be properly
quantified.

\medskip

Let us now turn to the error term. We want to approximate a column
vector of $C$ or $D$,
\begin{equation}
\label{eq:defu}
\mathbf{u}=\frac{1}{\sqrt{6}}(1,e^{2i\pi\phi_1},e^{2i\pi\phi_2},e^{2i\pi\phi_3},e^{2i\pi\phi_4},
e^{2i\pi\phi_5})
\end{equation}
by a vector $\tilde{\mathbf{u}}$ of the form \eqref{eq:approx}
where each $\phi_k$ has been approximated by some
$(j_k+1/2)/N'=r_k^{(N')}$. We must keep in mind that $\phi_k$ can
lie anywhere in the {\it interval} $[j_k/N', (j_k+1)/N')$. Let us
denote by $\mathbf{f}_0,\ldots,\mathbf{f}_5$ the columns of
$F(a,b)$ and by $\tilde{\mathbf{f}}_0,\ldots,\tilde{\mathbf{f}}_5$
those of $F(\tilde a,\tilde b)$. By construction
$\tilde{\mathbf{u}}$ and
$\tilde{\mathbf{f}}_0,\ldots,\tilde{\mathbf{f}}_5$ have the {\it
following property}: there exist numbers $\phi_k$ in $[j_k/N',
(j_k+1)/N')$, and numbers $a,b$ in $[\tilde a-\frac{1}{2N}, \tilde
a+\frac{1}{2N})$, $[\tilde b-\frac{1}{2N}, \tilde b+\frac{1}{2N})$
such that the corresponding vectors $\mathbf{u}$ (as in
\eqref{eq:defu}) and $\mathbf{f}_0, \dots \mathbf{f}_5$ (as in the
columns of \eqref{fab}) are unbiased to each other, i.e. $|\langle
\mathbf{u}, \mathbf{f}_k \rangle |=1/\sqrt{6}$. For a fixed pair
of discretization parameters $N,N'$ and a fixed pair of $\tilde a,
\tilde b$ the vectors $\tilde{\mathbf{u}}$ of the form
\eqref{eq:approx} which have the above property will be called
{\it quasi-unbiased to} $F(\tilde a, \tilde b)$, and their set
will be denoted by $FUB_{N,N'}^{\tilde a,\tilde b}$. Our first aim
is to find the vectors belonging to $FUB_{N,N'}^{\tilde a,\tilde
b}$ out of all vectors $\tilde{\mathbf{u}}$ of the form
\eqref{eq:approx}. Despite having $N'^5$ possible vectors
$\tilde{\mathbf{u}}$, we will see that the set $FUB_{N,N'}^{\tilde
a,\tilde b}$ will be of reasonably small size.

We will need the following lemma which we will refer to as the
{\it trivial error bound}.

\begin{lemma}\label{trivbound}
Let $I_k$ and $J_k$ ($1\le k\le 5)$ be closed intervals (possibly
degenerate) contained in $[0, 1]$. Let $L_k$ and $T_k$ denote the
lengths, while $m_k$ and $s_k$ the midpoints of the intervals
$I_k$ and $J_k$, respectively.

Consider the midpoint sum $S=\frac{1}{6}(1+\sum_{k=1}^5 e^{2i\pi
(m_k-s_k)})$. The following two statements hold:
\begin{itemize}
\item if it is possible to select points $\phi_k$ and $\psi_k$
from the intervals $I_k$ and $J_k$, such that
$|\frac{1}{6}(1+\sum_{k=1}^5 e^{2i\pi
(\phi_k-\psi_k)})|=\frac{1}{\sqrt{6}}$, then
\begin{equation}\label{trivunb} \frac{\pi}{6} \sum_{k=1}^{5}(L_k+T_k)\ge \left| |S|-\frac{1}{\sqrt{6}} \right|,\end{equation}

\item if it is possible to select points $\phi_k$ and $\psi_k$
from the intervals $I_k$ and $J_k$, such that
$\frac{1}{6}(1+\sum_{k=1}^5 e^{2i\pi (\phi_k-\psi_k)})=0$, then
\begin{equation}\label{trivort} \frac{\pi}{6} \sum_{k=1}^{5}(L_k+T_k)\ge|S|,\end{equation}
\end{itemize}
\end{lemma}

\begin{proof}
Let us introduce the ``error function" $E(\underline{x},
\underline{y})= S-\frac{1}{6}(1+\sum_{k=1}^5 e^{2i\pi
(x_k-y_k)}),$ where $x_k, y_k$ are in $I_k$ and $J_k$,
respectively. Note that $|(m_k-s_k)-(x_k-y_k)|\le
\frac{1}{2}(L_k+T_k)$ for each $1\le k\le 5$. The trivial estimate
$|e^{2i\pi (m_k-s_k)}-e^{2i\pi (x_k-y_k)}|\le \pi(L_k+T_k)$ yields
$|E(\underline{x}, \underline{y})|\le
\frac{\pi}{6}\sum_{k=1}^5(L_k+T_k).$ Therefore, the values of the
function $\frac{1}{6}(1+\sum_{k=1}^5 e^{2i\pi (\phi_k-\psi_k)})$
stay within a disk of radius $\frac{\pi}{6}\sum_{k=1}^5(L_k+T_k)$
around $S$. In the first statement of the lemma we thus conclude
that the distance of $S$ from the circle of radius $1/\sqrt{6}$
(centered at the origin) is not greater than $\frac{\pi}{6}
\sum_{k=1}^{5}(L_k+T_k)$. In the second statement we conclude that
the distance of $S$ from the origin is not greater than
$\frac{\pi}{6} \sum_{k=1}^{5}(L_k+T_k)$. These are equivalent to
\eqref{trivunb} and \eqref{trivort}.
\end{proof}

Let us apply this lemma to our particular case. The last 5
coordinates of $\tilde{\mathbf{u}}$ represent {\it intervals} of
length $1/N'$. For $j=0,2,4$ the columns $\tilde{\mathbf{f}}_j$ do
not contain the parameters $a,b$ so that all coordinates are known
{\it exactly}, which means that the corresponding intervals are
degenerate (of length zero). For $j=1,3,5$ the columns
$\tilde{\mathbf{f}}_j$ contain parameters at four different
coordinates, each representing {\it intervals} of the type
$[\tilde a-\frac{1}{2N}, \tilde a+\frac{1}{2N})$ and $[\tilde
b-\frac{1}{2N}, \tilde b+\frac{1}{2N})$, of length $1/N$.
Therefore Lemma \ref{trivbound} yields
\begin{equation}
\label{eq:bound1}
\abs{|\langle\tilde{\mathbf{u}},\tilde{\mathbf{f}}_j\rangle|-\frac{1}{\sqrt{6}}}
\leq\frac{5\pi}{6N'} \qquad j=0,2,4.
\end{equation}
\begin{equation}
\label{eq:bound2}
\abs{|\langle\tilde{\mathbf{u}},\tilde{\mathbf{f}}_j\rangle|-\frac{1}{\sqrt{6}}}
\leq \frac{5\pi}{6N'}+\frac{4\pi}{6N}\qquad j=1,3,5.
\end{equation}

However, these bounds turn out to be too crude, and we will need
the following {\it improved error bound}. The technical lemma
below establishes the simple fact that ``maximal error always
occurs at the endpoints of the intervals''.

\begin{lemma}\label{improved}
Let $I_k$ and $J_k$ ($1\le k\le 5)$ be closed intervals (possibly
degenerate) contained in $[0, 1]$. Let $L_k$ and $T_k$ denote the
lengths, $m_k$ and $s_k$ the midpoints, and $i_k^-, i_k^+$ and
$j_k^-, j_k^+$ the endpoints of the intervals $I_k$ and $J_k$,
respectively. Assume that $\sum_{k=1}^{5}(L_k+T_k)<\frac{1}{\pi}$.

\medskip

Consider the midpoint sum $S=\frac{1}{6}(1+\sum_{k=1}^5 e^{2i\pi
(m_k-s_k)})$, and all the 32 endpoint sums
$S_{\underline\epsilon}=\frac{1}{6}(1+\sum_{k=1}^5 e^{2i\pi
(i_k^{\epsilon_k}-j_k^{-\epsilon_k})})$ (where
$\underline\epsilon$ denotes any vector of $\pm$ signs; note that
$-\epsilon_k$ appears at the upper index of $j_k$). The following
two statements hold:
\begin{itemize}
\item if it is possible to select points $\phi_k$ and $\psi_k$
from the intervals $I_k$ and $J_k$, such that
$|\frac{1}{6}(1+\sum_{k=1}^5 e^{2i\pi
(\phi_k-\psi_k)})|=\frac{1}{\sqrt{6}}$, then
\begin{equation}\label{impunb} \max \{|S-S_{\underline\epsilon}|\}\ge||S|-\frac{1}{\sqrt{6}}|,\end{equation}

\item if it is possible to select points $\phi_k$ and $\psi_k$
from the intervals $I_k$ and $J_k$, such that
$\frac{1}{6}(1+\sum_{k=1}^5 e^{2i\pi (\phi_k-\psi_k)})=0$, then
\begin{equation}\label{import}\max \{|S-S_{\underline\epsilon}|\}\ge|S|,\end{equation}
\end{itemize}
\end{lemma}

\begin{proof}
Let $r= \max \{|S-S_{\underline\epsilon}|\}.$ Let us use again the
``error function" $E(\underline{x}, \underline{y})=
S-\frac{1}{6}(1+\sum_{k=1}^5 e^{2i\pi (x_k-y_k)}),$ where $x_k,
y_k$ are in $I_k$ and $J_k$, respectively. Apply for each $1\le
k\le 5$ the trivial estimate $|e^{2i\pi (m_k-s_k)}-e^{2i\pi
(x_k-y_k)}|\le \pi(L_k+T_k)$ to obtain $|E(\underline{x},
\underline{y})|\le
\frac{1}{6}\pi\sum_{k=1}^5(L_k+T_k)<\frac{1}{6},$ by assumption.
Also, $|E(\underline{x}, \underline{y})|$ is a continuous function
on a compact space, so it achieves its maximum.  We claim that the
maximum is achieved where all coordinates $x_k, y_k$ are opposite
endpoints of $I_k$ and $J_k$ (i.e. if $x_k$ is the lower endpoint
of $I_k$ then $y_k$ is the upper endpoint of $J_k$, and vice
versa). Assume by contradiction that this is not so for, say,
$x_1, y_1$. Then $x_1-y_1$ lies in the {\it interior} of the
interval $I_1-J_1$. This means that for $t$ small enough we can
move $x_1$ and/or $y_1$  to $x_1', y_1'$ within the intervals
$I_1, J_1$ so that $x_1'-y_1'=x_1-y_1+t$. This yields
\begin{eqnarray}|E(\underline{x'}, \underline{y'})|=|S-\frac{1}{6}-\frac{1}{6}e^{2i\pi
(x_1-y_1+t)}-\frac{1}{6}\sum_{k=2}^5 e^{2i\pi
(x_k-y_k)}|=\\
=|E(\underline{x}, \underline{y})+\frac{1}{6}(1-e^{
2ti\pi})e^{2i\pi (x_1-y_1)}|.\end{eqnarray} As $t$ varies in the
neighbourhood of zero, the locus of the points $E(\underline{x},
\underline{y})+\frac{1}{6}(e^{2ti\pi}-1)e^{2i\pi (x_1-y_1)}$ is a
small arc of a circle of radius $\frac{1}{6}$ with center
$E(\underline{x}, \underline{y})-\frac{1}{6}e^{2i\pi (x_1-y_1)} $.
This arc goes through $E(\underline{x}, \underline{y})$ at $t=0$.
Combining this with the fact that $|E(\underline{x},
\underline{y})|<\frac{1}{6}$, it results from easy plane geometry
that one can move along this circle in one way or the other so
that $|E(\underline{x'}, \underline{y'})|$ becomes larger than
$|E(\underline{x}, \underline{y})|$. The same argument applies to
any of the variables $x_k, y_k$, so we conclude that
$|E(\underline{x}, \underline{y})|$ indeed achieves its maximum
when all $x_k, y_k$ are at some opposite endpoints of the
intervals $I_k$ and $J_k$. This means that $r$ is the maximum of
$|E(\underline{x}, \underline{y})|$. Therefore, the values of the
function $\frac{1}{6}(1+\sum_{k=1}^5 e^{2i\pi (\phi_k-\psi_k)})$
stay within a disk of radius $r$ around $S$. In the first
statement of the lemma we thus conclude that the distance of $S$
from the circle of radius $1/\sqrt{6}$ (centered at the origin) is
not greater than $r$. In the second statement we conclude that the
distance of $S$ from the origin is not greater than $r$. These are
equivalent to \eqref{impunb} and \eqref{import}.
\end{proof}

We are now ready to search for vectors $\tilde{\mathbf{u}}\in
FUB_{N,N'}^{\tilde a,\tilde b}$. For each fixed pair\footnote{We
actually only take those $\tilde a,\tilde b$ which approximate
$a,b$ in the triangle $(0,0)$ $(1/6,0)$, $(1/6,1/12)$ {\it i.e.}
in the fundamental domain.} $\tilde a,\tilde
b=1/2N,3/2N,\ldots,(2N-1)/2N$ we simply take all possible values
of $j_1,\ldots,j_5$ from $0$ to $N'-1$ and check if the vector
$\tilde{\mathbf{u}}$ given by \eqref{eq:approx} satisfies the
bound \eqref{impunb} for all $\tilde{\mathbf{f}}_j$, $(0\le j\le
5).$ Recall that the coordinates of $\tilde{\mathbf{u}}$ and
$\tilde{\mathbf{f}}_j$ represent {\it intervals} in which the
actual coordinates of ${\mathbf{u}}$ and ${\mathbf{f}}_j$ might
lie. The sum of the lengths of these intervals is either
$\frac{5}{N'}$ or $\frac{5}{N'}+\frac{4}{N}$ (see equations
\eqref{eq:bound1},\eqref{eq:bound2}), which are less than
$\frac{1}{\pi}$ due to our choices $N=180$, $N'=19$, so that Lemma
\ref{improved} can indeed be applied.

It turns out, however, that the number of vectors
$\tilde{\mathbf{u}}$ satisfying the improved error bound
\eqref{impunb} is still too high.  Therefore we need to use the
following \emph{multiscale} strategy. We subdivide each interval
$I_j^{(N')}$ into two equal subintervals
$I_{j,-}^{(N')}=[r_j^{(N')}-1/(2N'),r_j^{(N')})$ and
$I_{j,+}^{(N')}=[r_j^{(N')},r_j^{(N')}+1/(2N'))$. Let $r_{j,-}$
and $r_{j,+}$ denote the midpoints of these subintervals. Clearly,
each $\phi_k$ --- defined in \eqref{eq:defu} --- must fall into one of
these intervals. This means that we can better approximate
$\mathbf{u}$ by
$$
\tilde{\mathbf{u}}_{\underline\epsilon}=\frac{1}{\sqrt{6}}(1,e^{2i\pi
r_{j_1,\epsilon_1}},\ldots,e^{2i\pi r_{j_5,\epsilon_5}}),
$$
where $\underline\epsilon$ is a $\pm$ vector with the signs being
chosen according to which subintervals $\phi_k$ fall. Then
$\tilde{\mathbf{u}}_{\underline\epsilon}$ is a better
approximation of $\mathbf{u}$ and needs to satisfy \eqref{impunb}
for all $\tilde{\mathbf{f}}_j$, $(0\le j\le 5)$, with the \emph{
smaller} intervals corresponding to
$\tilde{\mathbf{u}}_{\underline\epsilon}$. The $2^5$ vectors
$\tilde{\mathbf{u}}_{\underline\epsilon}$ will be called the
\emph{daughters} of $\tilde{\mathbf{u}}$ (called their
\emph{mother}). Clearly, if none of the daughters satisfies the
bound \eqref{impunb} then we can discard the mother. The point is
that it often happens that the mother satisfies the bound
\eqref{impunb} (corresponding to her own intervals), but none of
her daughters do. In such a situation we must keep the mother at
the first level of checking, but can discard her at the level of
daughters. We then repeat this operation, obtaining grandchildren
which have to satisfy the bound \eqref{impunb} for all
$\tilde{\mathbf{f}}_j$, $(0\le j\le 5)$, with even smaller
intervals. Again, if none of the grandchildren satisfies this
bound, then the grandmother is discarded. We repeat this operation
for 7 generations, i.e. a vector $\tilde{\mathbf{u}}$ of the form
\eqref{eq:approx} survives this test if and only if it has a
surviving descendant down to 7 generations. Our computer search
then shows that for a fixed pair of values $(\tilde a, \tilde b)$
we typically obtain 110-140 such vectors $\tilde{\mathbf{u}}\in
FUB_{N,N'}^{\tilde a,\tilde b}$. (This is quite satisfying
considering that $a,b$ are allowed to range over small intervals,
and we conjecture that the {\it precise} number of unbiased
vectors for any {\it exact} pair $a,b$ is 48). These vectors are
the candidates for the columns of $\tilde C, \tilde D$.

\medskip

Next, we attempt to compile the basis $\tilde C$. If there exists
a quartet $(Id,F(a,b),C,D)$ of mutually unbiased bases, then all
the columns of $\tilde C$ must come from the set
$FUB_{N,N'}^{\tilde a,\tilde b}$ and, furthermore, they must be
``almost-orthogonal" to each other. To be more precise, let
$\mathbf{u}=\frac{1}{\sqrt{6}}(1,e^{2i\pi\phi_1},\ldots,e^{2i\pi\phi_5})$
and
$\mathbf{v}=\frac{1}{\sqrt{6}}(1,e^{2i\pi\psi_1},\ldots,e^{2i\pi\psi_5})$
be any two vectors from $C$ and let
\begin{equation}\label{uhullam}\tilde{\mathbf{u}}=\frac{1}{\sqrt{6}}(1,e^{2i\pi
r_{j_1}^{(N')}},\ldots,e^{2i\pi r_{j_5}^{(N')}})\end{equation} and
\begin{equation}\label{vhullam}\tilde{\mathbf{v}}=\frac{1}{\sqrt{6}}(1,e^{2i\pi
r_{m_1}^{(N')}},\ldots,e^{2i\pi r_{m_5}^{(N')}})\end{equation} be
their approximation in $FUB_{N,N'}^{\tilde a,\tilde b}$. Recall
that we only have at hand the vectors $\tilde{\mathbf{u}},
\tilde{\mathbf{v}}$ and keep in mind that the actual phases of
$\mathbf{u}, \mathbf{v}$ can lie anywhere in the {\it intervals}
$I_{j_1}^{N'}, \dots, I_{j_5}^{N'}$ and $I_{m_1}^{N'}, \dots,
I_{m_5}^{N'}$, respectively. The fact that $\langle
\mathbf{u},\mathbf{v}\rangle =0$ implies the following condition
on $\tilde{\mathbf{u}}, \tilde{\mathbf{v}}$:
\begin{definition} We will say that the vectors
$\tilde{\mathbf{u}}$ and $\tilde{\mathbf{v}}$ of the form
\eqref{uhullam}, \eqref{vhullam} are {\em $N'$-orthogonal} if
there exist numbers $\phi_k$ and $\psi_k$ in the intervals
$I_{j_k}^{(N')}$ and $I_{m_k}^{(N')}$, such that
$\frac{1}{6}(1+\sum_{k=1}^5 e^{2i\pi (\phi_k-\psi_k)})=0$.

Similarly, we will say that $\tilde{\mathbf{u}}$ and
$\tilde{\mathbf{v}}$ are {\it $N'$-unbiased} if there exist
$\phi_k\in I_{j_k}^{(N')}$ and $\psi_k\in I_{m_k}^{(N')}$, such
that $|\frac{1}{6}(1+\sum_{k=1}^5 e^{2i\pi
(\phi_k-\psi_k)})|=\frac{1}{\sqrt{6}}$.
\end{definition}
These properties are clearly rotation invariant in the sense that
they only depend on the values $e^{2i\pi
(r_{j_1}^{(N')}-r_{m_1}^{(N')})},\ldots, e^{2i\pi
(r_{j_5}^{(N')}-r_{m_5}^{(N')})}$, i.e. the values of
$r_{j_1}^{(N')}-r_{m_1}^{(N')}, \ldots,
r_{j_5}^{(N')}-r_{m_5}^{(N')}$ modulo 1. We can therefore take
$m_1=\dots=m_5=0$ and correspondingly
$\tilde{\mathbf{v}}_0=\frac{1}{\sqrt{6}}(1,e^{i\pi
/N'},\ldots,e^{i\pi/N'})$, (where the exponents of the last 5
coordinates represent intervals, of course) and define the set
$ORT_{eps, N'}$ as the set of vectors which are $N'$-orthogonal to
$\tilde{\mathbf{v}}_0$. With this notation the rotation invariance
means that $\tilde{\mathbf{u}}$ and $\tilde{\mathbf{v}}$ as in
\eqref{uhullam}, \eqref{vhullam} are $N'$-orthogonal if and only
if the vector $\frac{1}{\sqrt{6}}(1,e^{2i\pi
(r_{j_1}^{(N')}-r_{m_1}^{(N')})},\ldots, e^{2i\pi
(r_{j_5}^{(N')}-r_{m_5}^{(N')})})$ is in $ORT_{eps, N'}$.

\medskip

Note that the set $ORT_{eps, N'}$ is independent of $\tilde a,
\tilde b$, so it can be computed once and for all at the beginning
of the computer search. In order to find the set $ORT_{eps, N'}$
we first introduce the simpler set $ORT_{N'}$ as the set of
vectors $\tilde{\mathbf{u}}$ (as in \eqref{uhullam}) for which
there exist $\phi_k\in I_{j_k}$ such that
$\frac{1}{6}(1+\sum_{k=1}^5 e^{2i\pi \phi_k})=0$. We will check
each possible vector $\tilde{\mathbf{u}}$ whether it is in
$ORT_{N'}$ (recall that there are $N'^5$ possibilities for
$\tilde{\mathbf{u}}$). In order to do so, we apply Lemma
\ref{improved} to $\tilde{\mathbf{u}}$ and the {\it exact} vector
${\mathbf{v}}_0=(1,1, \dots 1)$ (so that in the notations of the
Lemma the intervals $J_k$ are degenerate). We then use our
multiscale strategy again, i.e. we test the descendants of
$\tilde{\mathbf{u}}$ against ${\mathbf{v}}_0$ with Lemma
\ref{improved} down to 7 generations. We keep only those vectors
$\tilde{\mathbf{u}}$ which have at least one surviving descendant
in each generation. Having constructed the set $ORT_{N'}$ it is
now easy to obtain $ORT_{eps, N'}$. Indeed, by definition a vector
$\tilde{\mathbf{u}}=\frac{1}{\sqrt{6}}(1,e^{2i\pi
r_{j_1}^{(N')}},\ldots,e^{2i\pi r_{j_5}^{(N')}})$ can only be
$N'$-orthogonal to $\tilde{\mathbf{v}}_0$ if there exist numbers
$\phi_k$ in the intervals $I_{j_k}$ and $\psi_k$ in
$[0,\frac{1}{N'})$, such that $\frac{1}{6}(1+\sum_{k=1}^5 e^{2i\pi
(\phi_k-\psi_k)})=0$. But then the numbers $\phi_k-\psi_k$ must
fall in the intervals $I_{j_k-\epsilon_k}$ where $\epsilon_k$ is
either 0 or 1, and hence the vector
$\tilde{\mathbf{u}}_\epsilon=\frac{1}{\sqrt{6}}(1,e^{2i\pi
r_{j_1-\epsilon_1}^{(N')}},\ldots,e^{2i\pi
r_{j_5-\epsilon_5}^{(N')}})$ is in $ORT_{N'}$. Therefore, to
construct $ORT_{eps, N'}$ we take all vectors of the form
$\tilde{\mathbf{u}}^\epsilon=\frac{1}{\sqrt{6}}(1,e^{2i\pi
r_{j_1+\epsilon_1}^{(N')}},\ldots,e^{2i\pi
r_{j_5+\epsilon_5}^{(N')}})$, where $\epsilon_k$ is 0 or 1, and
the vector $\frac{1}{\sqrt{6}}(1,e^{2i\pi
r_{j_1}^{(N')}},\ldots,e^{2i\pi r_{j_5}^{(N')}})$ is in
$ORT_{N'}.$ In the specific case $N'=19$ we found that the set
$ORT_{eps, N'}$ contains $322040$ vectors. This means that the
``probability" of two random vectors being $N'$-orthogonal is
$322040/19^5\approx 0.13.$

\medskip

Having constructed the set $ORT_{eps, N'}$ we search for the
columns of the matrix $\tilde C$ in such a way that for any two
columns $\tilde{\mathbf{u}}$ and $\tilde{\mathbf{v}}$ (as in
\eqref{uhullam}, \eqref{vhullam}) we require that
$\tilde{\mathbf{u}}, \tilde{\mathbf{v}}\in FUB_{N,N'}^{\tilde
a,\tilde b}$ and that the vector $\frac{1}{\sqrt{6}}(1,e^{2i\pi
(r_{j_1}^{(N')}-r_{m_1}^{(N')})},\ldots, e^{2i\pi
(r_{j_5}^{(N')}-r_{m_5}^{(N')})})$ be in $ORT_{eps, N'}$.

\medskip

Let us make a last simplifying remark. It is clear that we can
permute the columns of all appearing matrices, and hence we are
free to choose the {\it order} of the columns of $\tilde C$.
Therefore we assume in our search that the columns of $\tilde C$
are lexicographically ordered, meaning that for any two columns
$\tilde{\mathbf{u}}$ and $\tilde{\mathbf{v}}$ (as in
\eqref{uhullam}, \eqref{vhullam}) we have that
$\tilde{\mathbf{u}}$ precedes $\tilde{\mathbf{v}}$ if and only if
$(j_1, \dots, j_5)$ precedes $(m_1, \dots, m_5)$ in lexicographic
order.

\medskip

Our computer search has shown that for a fixed pair of values
$(\tilde a, \tilde b)$ there are typically 1000-5000 such
$N'$-orthonormal bases $\tilde C$. (At some special values of
$(\tilde a, \tilde b)$, however, there are millions. (This nicely
complies with the finding of \cite{ujbrit} that for $(a,b)=(1/6,
0)$ there exist 70 bases $C$, whereas for generic values of
$(a,b)$ there exist only 8.)

\medskip

Finally, for any fixed pair $(\tilde a, \tilde b)$ and any
corresponding matrix $\tilde C$ we attempt to compile the basis
$\tilde D$. The columns of $\tilde D$ must also come from the set
$FUB_{N,N'}^{\tilde a,\tilde b}$, they must be $N'$-orthogonal to
each other, and they must be $N'$-unbiased to the columns of
$\tilde C$. Therefore, to find the candidates for the columns of
$\tilde D$ we will check any vector $\tilde{\mathbf{u}}\in
FUB_{N,N'}^{\tilde a,\tilde b}$ whether it is $N'$-unbiased
simultaneously to all the columns $\tilde{\mathbf{c}}_k$ of
$\tilde C$. This is done by applying the trivial bound, Lemma
\ref{trivbound}, to the vector $\tilde{\mathbf{u}}$ (and its
descendants for 7 generations) and the vectors
$\tilde{\mathbf{c}}_k$ for $k=0, \dots, 5$. The reason why we use
the trivial bound instead of the improved bound is that it speeds
up calculations and very few vectors $\tilde{\mathbf{u}}$ survive
this test anyway. Let $COL_{\tilde{D}}$ denote the set of
surviving vectors, the candidates for the columns of $\tilde D$.
If there are less than 6 vectors in $COL_{\tilde{D}}$ then we
conclude that $\tilde D$ cannot exist (as it would need 6
columns). If there are at least 6 vectors in $COL_{\tilde{D}}$
then we check whether any 6 of them can be pairwise
$N'$-orthogonal to each other (this is done by using the set
$ORT_{eps, N'}$ again).

Our computer search shows that there are no values of $(\tilde a,
\tilde b)$ and corresponding $\tilde C$ for which all these
conditions on $\tilde D$ can be met. This concludes the proof of
the theorem.
\end{proof}

\bigskip

\begin{remark}\rm
In {\it principle}, this discretization scheme could successfully
be applied to settle Problem \ref{MUB6problem}. Of course, in the
general case we cannot assume that $B$ is of the form $B=F(a,b)$.
However, we know that $B$ is {\it some} complex Hadamard matrix.
If a complete classification of complex Hadamard matrices of order
6 were available in some parametric form then a similar search
could be carried out as above. Without such classification at hand
we can still use a {\it finite} set of $N$ representatives in {\it
each coordinate} of $B$ to approximate it with a quasi-Hadamard
matrix $\tilde{B}$. The rest of the algorithm concerning the
selection of quasi-unbiased vectors, and the checking of the
possibly arising matrices $\tilde{C}$ and $\tilde{D}$ remains the
same. Note that $B$ has 25 free entries (as the first row and
column can be assumed to be 1). At first glance an exhaustive
search for $B$ should go through $N^{25}$ cases, way out of the
realm of possibilities, if $N\approx 100$. However, one can reduce
the number of cases with intelligent tricks, so that the search
can actually be carried out. The problem is that while in the case
of $B=F(a,b)$ and $N=180$ we had only 270 candidates for
$\tilde{B}$, in the general case we have about $10^{12}$ such
candidates already at $N\approx 100$. And for each candidate
$\tilde{B}$ we need to run the final part of the algorithm
concerning the selection of unbiased vectors and compiling the
bases $\tilde{C},\tilde{D}$. While it is not absolutely out of the
question to carry out a computer search at such magnitudes, it
definitely needs meticulous programming and probably some further
mathematical ideas to reduce the number of cases.

Let us also recall that the non-existence of projective planes of
order 10 was also shown by an exhaustive computer search
\cite{lam} --- and no ``theoretical" proofs are known. 
\end{remark}

\section{Smooth families of mutually unbiased bases}
\label{sec:theory}

Throughout this section we will assume that
\begin{equation}
{\mathcal F}(t) = (\mathbf{f}_1(t),\ldots \mathbf{f}_d(t))
\end{equation}
is a ``time-dependent'' family of orthonormal bases. More
precisely, we assume that the maps $I\ni t \mapsto \mathbf
f_k(t)\in \C^d$  (for $k=1,\ldots,d$) are smooth (where $I\subset
\mathbb R$ is a certain fixed open interval) and that
$\mathcal{F}(t)$ is an orthonormal basis (ONB) for each $t\in I$.
We shall then say that $t\mapsto ({\mathcal E},{\mathcal F}(t))$
is a {\bf smooth family of pairs of MUB} if ${\mathcal E} =
(\mathbf{e}_1,\ldots \mathbf{e}_d)$ is a fixed ONB such that
${\mathcal E}$ and ${\mathcal F}(t)$ are mutually unbiased for all
$t\in I$.

\subsection{Smooth families of MUB and common unbiased vectors}

Let us consider now the following question. Assume we are given a
vector $\mathbf{b}_0$ of unit length which is unbiased to both
${\mathcal E}$ and ${\mathcal F}(0)$. Can we ``continue''
$\mathbf{b}_0$ so as to find a common unbiased vector
$\mathbf{b}(t)$ for ${\mathcal E}$ and ${\mathcal F}(t)$ for $t$
in a neighbourhood of $0$?

In order to answer this question we shall need some further
notions and notations. First, if $\mathbf{u}$ is any vector in
$\C^d$ with $\norm{\mathbf{u}}=1$, let
$\mathcal{P}_{\mathbf{u}}$ be the ortho-projection on the span of
$\mathbf{u}$; {\it i.e.} $\mathcal{P}_{\mathbf{u}}\mathbf
x=\scal{\mathbf x,\mathbf{u}}\mathbf{u}$. Next, consider the
$d\times d$ complex matrix $M=[M_{k,l}]_{1\le k,l\le d}$ defined
by the formula
\begin{equation}
\label{defofM} M_{k,l}:={\rm Tr}(P_kQ_lR)
\end{equation}
where $P_k=\mathcal{P}_{\mathbf{e}_k}$,
$Q_l=\mathcal{P}_{\mathbf{f}_l(0)}$ and
$R=\mathcal{P}_{\mathbf{b}_0}$. A simple computation shows that
$$
M_{k,l} =
 \langle \mathbf{b}_0, \mathbf{e}_k \rangle \, \langle
\mathbf {e}_k , \mathbf{f}_l(0) \rangle \, \langle \mathbf{f}_l(0)
, \mathbf{b}_0 \rangle.
$$
Note that $M_{k,l}$ is unchanged if any of the vectors
$\mathbf{b}_0, \mathbf{e}_k,\mathbf{f}_l(0)$ is multiplied by a
complex number of modulus $1$. Moreover, $M=D_1HD_2$ where $D_1$
is the non-singular diagonal matrix with entries $\langle
\mathbf{b}_0, \mathbf{e}_k \rangle$, $D_2$ is the non-singular
diagonal matrix with entries $\langle \mathbf{f}_l(0) ,
\mathbf{b}_0 \rangle$ and $H=[\langle \mathbf {e}_k ,
\mathbf{f}_l(0) \rangle]_{1\le k,l\le d}$ is a multiple of a
Hadamard matrix. Thus $M$ is of rank $d$. However, the {\it real}
matrix $N$ whose entries are obtained by taking the imaginary part
of the entries of $M$,
\begin{equation}\label{defofN}
N_{k,l}:= {\rm Im}(M_{k,l}) = \frac{1}{2i} {\rm Tr}([P_k,Q_l]R)
\end{equation}
cannot be of maximal rank. Indeed,  for all $l=1,\ldots, d$,
$$
\sum_k N_{k,l} = \sum_k \frac{1}{2i} {\rm Tr}([P_k,Q_l]R) =
\frac{1}{2i} {\rm Tr}([\mathbf 1,Q_l]R) = 0
$$
since $\sum_k P_k =\mathbf 1$.

\begin{definition}
We say that a unit vector $\mathbf{b}$ that is unbiased to both
members of a MUB pair $(\mathcal{E},\mathcal{F})$ is
\emph{non-degenerate} if the associated real matrix $N(\mathbf b)
= [{\rm Im}(\langle \mathbf{b}, \mathbf{e}_k \rangle \langle
\mathbf {e}_k , \mathbf{f}_l \rangle \langle \mathbf{f}_l ,
\mathbf{b} \rangle)]_{1\le k,l\le d}$ has rank $d-1$.
\end{definition}

\begin{theorem}\label{UBcontinuity}
Let $({\mathcal E},{\mathcal F}(t))$ be a smooth family of pairs
of MUB. Let $\mathbf b_0$ be a non-degenerate common normalized
unbiased vector for the MUB pair $(\mathcal E,\mathcal F(0))$.
Then there exists an $\epsilon > 0$ and a smooth map
$(-\epsilon,\epsilon)\ni t \mapsto \mathbf b(t)$ such that
$\mathbf b(0)=\mathbf b_0$ and $\mathbf b(t)$ is a common
normalized unbiased vector to $(\mathcal E,\mathcal F(t))$ for all
$t\in (-\epsilon,\epsilon)$.
\end{theorem}

\begin{proof}
We may modify the order of columns and rows of $N:=N({\mathbf
b_0})$ by re-ordering the vectors in $\mathcal E$ and ${\mathcal
F}$. But, from the rank condition, $N$ has a $(d-1)\times(d-1)$
invertible submatrix so, without loss of generality, we may assume
that the $(d-1)\times(d-1)$ submatrix in the upper-left corner of
$N$ is invertible.

A vector $\mathbf v$ is unbiased to a given pair of MUB if and
only if $\lambda \mathbf v$ is so, where $\lambda \in \mathbb C,
|\lambda|=1$. Thus, without loss of generality, we may assume that
$\scal{\mathbf{e}_d,\mathbf b_0}=d^{-1/2}$ and in fact we may
further require $t\mapsto \mathbf b(t)$ to satisfy
$\scal{\mathbf{e}_d,\mathbf b(t)}=d^{-1/2}$ for all $t\in
(-\epsilon,\epsilon)$. (Indeed, if $t\mapsto \mathbf b(t)$
satisfies the original requirements of our theorem, then $t\mapsto
\tilde{\mathbf{b}}(t):= \alpha(t)\mathbf{b}(t)$ where $\alpha(t)
=d^{-1/2}\scal{\mathbf e_d, \mathbf{b}(t)}$ satisfies the just
introduced extra condition, too.) Then the condition
$\abs{\scal{\mathbf{e}_j,\mathbf b(t)}} =d^{-1/2}$ (for
$j=1,\ldots, d-1$), together with the condition of smoothness, are
equivalent to saying that $\mathbf{b}(t) =
\mathbf{v}\bigl(x(t)\bigr)$ where
\begin{equation}
\mathbf{v}(x) =d^{-1/2}\left( \sum_{j=1}^{d-1} e^{ix_j}
\mathbf{e}_j \,+\, \mathbf{e}_d\right)
\end{equation}
and $t\mapsto x(t)\in \mathbb R^{d-1}$ is a real smooth curve. In
this rephrasing of the problem the initial condition
$\mathbf{b}(0)=\mathbf{b}_0$ reads as $x(0)=x^{(0)}$ where
$x^{(0)}=(x^{(0)}_1,\ldots,x^{(0)}_{d-1})\in\mathbb R^{d-1}$ is such that
$\scal{\mathbf{b}_0,\mathbf{e}_j}=d^{-1/2}e^{ix^{(0)}_j}$. Note
also that $\norm{\mathbf{v}(x)}=1$ is automatically satisfied.

Let us now introduce the function $u: I \times \mathbb R^{d-1} \to
\mathbb R^{d-1}$ defined by the formula
\begin{equation}
\label{eq:defub0} u_k(t,x) =
\abs{\scal{\mathbf{f}_k(t),\mathbf{v}(x)}} - \frac{1}{\sqrt{d}} \
\ \ \ (k=1, \dots, d-1)
\end{equation}
and note that $\mathbf v(x)$ is unbiased to $\mathcal F(t)$ if and
only if $u(t,x)=0$; that is, if $u_k(t,x)=0$ for all
$k=1,...,d-1$.

By assumption, $u(0,x^{(0)})=0$ since $\mathbf v(x^{(0)})=\mathbf
b_0$ is an unbiased vector for $\mathcal F(0)$. Further, as
$|\langle \mathbf f_k, \mathbf b_0\rangle |=d^{-1/2}\neq 0$ for
all $k=1,..,d-1$, the function $u$ is smooth in a neighbourhood of
$(0,x^{(0)})$.\footnote{Actually, if $s\mapsto z(s)\in \mathbb C$
is smooth and $z(s)\neq 0$ then
$\dst\frac{\mathrm{d}}{\mathrm{d}s}|z(s)| = \frac{1}{|z(s)|}{\rm
Re}\,(\overline{z'(s)}z(s))$.} A simple computation then shows
$\partial_{x_k}\mathbf{v}(x) =  i \langle \mathbf{v}(x),
\mathbf{e}_k, \rangle \mathbf{e}_k$ form which we deduce
\begin{eqnarray*} \nonumber
\partial_{x_k} u_j(t,x) &=& \frac{1}{\abs{\scal{\mathbf{f}_j(t),
\mathbf{v}(x)}}} {\rm Re}\,\Bigl(
\scal{\partial_{x_k}\mathbf{v}(x),\mathbf{f}_j(t)}
\scal{\mathbf f_j(t),\mathbf{v}(x)}\Bigr)\\
&=&  \frac{1}{\abs{ \scal{\mathbf{f}_j(t),\mathbf{v}(x)}}} {\rm
Re}\,\Bigl(i \scal{\mathbf{v}(x),\mathbf{e}_k}
\scal{\mathbf{e}_k,\mathbf{f}_j(t)}
\scal{\mathbf{f}_j(t),\mathbf{v}(x)}\Bigr).
\end{eqnarray*}
Therefore
\begin{eqnarray*}\nonumber
\partial_{x_k} u_j(t,x)\Big|_{(t,x)=(0,x^{(0)})}&=&
\sqrt{d}\, {\rm Re}\,\Bigl(i \scal{\mathbf{b}_0,\mathbf{e}_k}
\scal{\mathbf{e}_k,\mathbf{f}_j}
\scal{\mathbf{f}_j,\mathbf{b}_0}\Bigr)\nonumber\\ \nonumber &=&-
\sqrt{d} \,{\rm Im}\,\Bigl( \scal{\mathbf{b}_0,\mathbf{e}_k}
\scal{\mathbf{e}_k,\mathbf{f}_j} \scal{\mathbf{f}_j,\mathbf{b}_0}\Bigr)\\
&=& -\sqrt{d} \,N_{k,j}.
\end{eqnarray*}
Thus the Jacobian of $u(0,\cdot)$ at $x=x^{(0)}$ is a nonzero
multiple of the $(d-1)\times(d-1)$ submatrix in the upper-left
corner of $N$, and the theorem follows by a use of the Implicit
Function Theorem.
\end{proof}

Note that the Implicit Function Theorem, used in the above proof,
actually tells us more than just existence. The invertibility of
the Jacobian of the function $u(0,\cdot):\ \mathbb
R^{d-1}\to\mathbb R^{d-1}$ at $x=x^{(0)}$ guarantees the existence
of a neighbourhood of $(0,x^{(0)})$ in which the only solution of
$u(t,x)=0$ is $(t,x)=(t,x(t))$. In particular, for $|t|$ small
enough, in a neighbourhood of $\mathbf{b}(t)$, the only vectors
that are unbiased for the MUB pair $(\mathcal E,\mathcal F(t))$
are the multiples of $\mathbf{b}(t)$.

\begin{corollary}\label{isolatedUB}
Let $\mathbf{b}$ be a common non-degenerate unbiased vector for
the MUB pair $\bigl(\mathcal{E},\mathcal{F}\bigr)$. Then there is
a neighbourhood of $\mathbf b$ in which all common unbiased
vectors to  $\bigl(\mathcal{E},\mathcal{F}\bigr)$ are multiples of
$\mathbf b$.
\end{corollary}

Actually, our theorem also allows to prove that the number of
vectors (counted up to multiples) that are unbiased to the family
$({\mathcal E},{\mathcal F}(t))$ is (under some non-degeneracy
conditions) independent of $t$ (for $|t|$ small enough).

\begin{corollary}\label{UBnumberunchanged}
Let $({\mathcal E},{\mathcal F}(t))$ be a smooth family of pairs
of MUB, and assume that every common normalized unbiased vector to
$({\mathcal E},{\mathcal F}(0))$ is non-degenerate. Then there
exists an $\epsilon>0$ such that for $|t|<\epsilon$, the number of
common normalized unbiased vectors to $({\mathcal E},{\mathcal
F}(t))$, when counted up to multiples, is finite and independent
of $t$. Moreover each of these vectors is given by Theorem
\ref{UBcontinuity}.
\end{corollary}

\begin{proof}
At $t=0$, Corollary \ref{isolatedUB} implies that each normalized
vector that is unbiased to both ${\mathcal E}$, and ${\mathcal
F}(0)$ is isolated in the explained sense. Since the unit sphere
of $\C^d$ is compact, it follows that up to multiples, there can
be only finitely many such vectors; say
$\mathbf{b}^{(1)},\ldots,\mathbf{b}_0^{(m)}.$ According to Theorem
\ref{UBcontinuity}, for each of these vectors, there is a smooth
curve $t\mapsto \mathbf b^{(k)}(t)$ such that $\mathbf
b^{(k)}(0)=\mathbf b^{(k)}_0$ and $\mathbf{b}^{(k)}(t)$ is
unbiased to $({\mathcal E},{\mathcal F}(t))$. By the comment
before Corollary \ref{isolatedUB} and by the fact that the just
introduced $m$ is a finite number, there exist some
$\tilde{\epsilon},r>0$ such that if $|t|<\tilde{\epsilon}$ then
none of the vectors $\mathbf{b}^{(1)}(t),\ldots,
\mathbf{b}^{(m)}(t)$ are multiples of each other and if $\mathbf
b$ is a common unbiased vector to $(\mathcal E,\mathcal F(t))$,
then for every $k=1,\ldots,m$, either
$$
\|\mathbf b - \mathbf{b}^{(k)}(t)\|>r
$$
or $\mathbf b$ is a multiple of $\mathbf{b}^{(k)}(t)$. In
particular, the number of of common normalized unbiased vectors to
$({\mathcal E},{\mathcal F}(t))$, counted up to multiples, is at
least $m$ (since we have the vectors $\mathbf{b}^{(1)}(t),\ldots,
\mathbf{b}^{(m)}(t)$.)

To prove the remaining part of our statement, we shall argue by
contradiction. Assume there is no such $\epsilon>0$ whose
existence is stated in our theorem. Then there should exist a real
sequence $t_n\; (n\in \mathbb N)$ converging to $0$ and a sequence
of unit vectors $\mathbf b_n \; (n\in \mathbb N)$ such that for
every $n\in \mathbb N$
\begin{itemize}
\item
$\mathbf b_n$ is a common unbiased vector to $(\mathcal E,
\mathcal F(t_n))$,
\item
$\mathbf b_n$ is not a multiple of any of the vectors
$\mathbf{b}^{(1)}(t_n),\ldots,\mathbf{b}^{(m)}(t_n)$.
\end{itemize}
Since the unit sphere of $\C^d$ is compact, it follows that there
is a subsequence of $\mathbf b_n \; (n\in \mathbb N)$ which is
convergent. In fact, without loss of generality we may assume that
our original sequence was such. Let $\mathbf b:={\rm lim}_n
(\mathbf b_n)$; it is then clear that $\|\mathbf b\|=1$ and since
$|\langle \mathbf b_n,\mathbf e_k \rangle| = |\langle \mathbf
b_n,\mathbf f_j(t_n) \rangle| = d^{-1/2}$, by continuity of the
scalar product, absolute value, and the map $t\mapsto \mathbf
f_j(t)$, we have that $\mathbf b$ is a common unbiased vector to
$(\mathcal E,\mathcal F(0))$. Hence by assumption there must exist
a $k$ such that $\mathbf b$ is a multiple of $\mathbf{b}^{(k)}_0$.
Actually it is clear, that we even may assume that $\mathbf b$ is
not only a {\it multiple} of $\mathbf{b}^{(k)}_0$, but equal to
it. We can then conclude our proof since as $n\to \infty$, we have
$$
r < \|\mathbf b_n - \mathbf b^{(k)(t_n)}\|  \to
\|\mathbf{b}^{(k)}_0-\mathbf{b}^{(k)}_0\| =0
$$
which is clearly a contradiction.
\end{proof}

\subsection{Unitary symmetries of mutually unbiased bases}

Recall that if ${\mathcal E} = (\mathbf{e}_1,\ldots \mathbf{e}_n)$
and ${\mathcal F} =(\mathbf{f}_1,\ldots \mathbf{f}_n) $ are two
mutually unbiased ONBs, then upon multiplying each vector by a
complex number of modulus 1 and changing the orders of the vectors
in the individual bases, they still remain two mutually unbiased
ONBs. In order not to distinguish between such pairs, we will
associate to an ONB ${\mathcal E} =(\mathbf{e}_1,\ldots
\mathbf{e}_d)$ a maximal abelian star algebra
$$
{\mathcal A}_{{\mathcal E}}:={\rm Span}\{\mathcal{P}_{\mathbf
e_k}| k=1,\ldots,d\}
$$
where $\mathcal{P}_{\mathbf e_k}$ is the ortho-projection onto
${\mathbb C}\mathbf e_k$, $(k=1,\ldots, d)$. Indeed, ${\mathcal
A}_{{\mathcal E}}$ is invariant under reordering and changing
phases of the vectors in ${\mathcal E}$. Moreover, as is well
known and easy to show, ${\mathcal E} = (\mathbf{e}_1,\ldots
\mathbf{e}_n)$ and ${\mathcal F} =(\mathbf{f}_1,\ldots
\mathbf{f}_n)$ are two mutually unbiased ONBs if and only if
${\mathcal A}_{{\mathcal E}}$ and ${\mathcal A}_{{\mathcal F}}$
are \emph{quasi-orthogonal}. Recall that this means that the
subspaces of $M_d({\mathbb C})$ given by  $\mathbf{1}^\perp
\cap{\mathcal A}_{{\mathcal E}}$ and $\mathbf{1}^\perp
\cap{\mathcal A}_{{\mathcal F}}$ are orthogonal with respect to
the usual Hilbert-Schmidt scalar product
$\scal{A,B}_{M_d(\C)}=\mbox{Tr}\,(A^*B)$ on $M_d({\mathbb C})$.

Further, to a unitary operator $U$ on $\mathbb{C}^d$, we associate
the authomorphism $\alpha_U$ defined by the formula
$\alpha_U(X)=UXU^*$. We will say that $U$ \emph{implements a
symmetry} of $(\mathcal E,\mathcal F)$ if $\alpha(\mathcal
A_{\mathcal E})=\mathcal A_{\mathcal E}$ and $\alpha(\mathcal
A_{\mathcal F})=\mathcal A_{\mathcal F}$. Accordingly, we shall
talk about the \emph{unitary symmetry group} of $({\mathcal E},
{\mathcal F})$.

There is a natural homomorphism from the group of symmetries of
$({\mathcal E}, {\mathcal F})$ to $S_d \times S_d$ where $S_d$ is
the group of permutations of $d$ elements. Indeed, a symmetry
takes a minimal projection of ${\mathcal A}_{{\mathcal E}}$ and
${\mathcal A}_{{\mathcal F}}$ into a minimal projection of
${\mathcal A}_{{\mathcal E}}$ and ${\mathcal A}_{{\mathcal F}}$,
respectively. Thus if $\alpha$ is a symmetry, then there exist two
permutations $\sigma=\sigma_\alpha,\mu=\mu_\alpha \in S_d$ such
that
$$
\alpha(P_{\mathbf e_k}) = P_{\mathbf e_{\sigma(k)}} \;\;\;\;
\textrm{and}\;\;\;\; \alpha(P_{\mathbf f_k}) = P_{\mathbf
f_{\mu(k)}}
$$
for all $k=1,\ldots,d$. Moreover, it is straightforward to show
that the map that associates the pair $(\sigma,\mu)$ to $\alpha$,
defines a group homomorphism.

\begin{theorem}
The homomorphism from the group of unitary symmetries to
$S_n\times S_n$ defined above is injective. In particular the
group of unitary symmetries can have at most $(d!)^2$ elements.
\end{theorem}

\begin{proof}
For the injectivity all we need to show is that if $U$ is a
unitary operator such that $UP_{\mathbf e_k}U^* = P_{\mathbf e_k}$
and $UP_{\mathbf f_k}U^* = P_{\mathbf f_k}$  for all $k=1,\ldots,
n$ then $U$ is a multiple of $\mathbf 1$.

However, the assumed invariance means that both the vectors of
${\mathcal E}$ and the vectors of ${\mathcal F}$ are eigenvectors
for $U$. Thus $U$ commutes with all elements of ${\mathcal
A}_{{\mathcal E}}$ and ${\mathcal A}_{{\mathcal F}}$. As both of
these are maximal abelian, it follows that $U\in {\mathcal
A}_{{\mathcal E}}\cap {\mathcal A}_{{\mathcal F}}$. However, by
the quasi-orthogonality this intersection contains only multiples
of the identity.
\end{proof}

Suppose  $({\mathcal E},{\mathcal F})$ is a MUB pair and that the
unitary operator $U$ implements a symmetry of $({\mathcal
E},{\mathcal F})$. Since $U$ may only reorder and multiply by unit
complex numbers the vectors of both $\mathcal E$ and $\mathcal F$,
it is clear that if $\mathbf b$ is a common unbiased vector to
$({\mathcal E},{\mathcal F})$, then so is $U\mathbf b$. Thus such
unitary operators allow us to construct (possibly new) common
unbiased vectors, once we have at least one such vector.

Before giving a general result, let us show how this may be used
to construct an ONB that is unbiased to both the standard basis
$\mathcal{E}$ and to the Fourier basis of $\mathbb C^d$. Let $U,V$
be the linear operators defined by the formulae
\begin{equation}\label{defUV}
U\mathbf e_k = e^{i\frac{2\pi}{d}(k-1)} \mathbf e_k,\;\;\;\;
\textrm{and} \;\;\;\; V\mathbf e_k = \mathbf e_{k-1}
\end{equation}
where the index ``$k-1$'' is meant by modulo $d$. Then $U,V$ are
unitary, $U^d=V^d=\mathbf 1$ and $VU = e^{i\frac{2\pi}{d}} UV$.
However, by the definition of the Fourier basis, a simple check
shows that
\begin{equation}\label{UVinF}
V\mathbf f_k = e^{i\frac{2\pi}{d}(k-1)} \mathbf f_k,\;\;\;\;
\textrm{and}\;\;\;\; U\mathbf f_k = \mathbf f_{k+1}
\end{equation}
for all $k=1,\ldots, d$ (where the index ``$k+1$'' is again meant
by modulo $d$). Thus $U$ and $V$ only change the ``phases'' and
reorder the vectors of both ${\mathcal E}$ and ${\mathcal F}$,
thus they implement unitary symmetries of $({\mathcal E},{\mathcal
F})$. In particular, if $\mathbf b$ is a common UB vector for both
the standard and the Fourier basis, then so is $U^kV^l\mathbf b$
for all $k,l=1,\ldots,d$.

As is well known in case of the Fourier basis, if $\mathbf b$ is a
common normalized UB vector for $({\mathcal E},{\mathcal F})$ then
the vectors
$$
\mathbf b, V\mathbf b, V^2 \mathbf b,\ldots , V^{d-1}\mathbf b
$$
form an ONB. The same stays true if one replaces $V$ by $U$.

We would like now to extend this to more general pairs of unbiased
bases. To do so, notice first that in the above case, the natural
injection from unitary symmetries into $S_d\times S_d$ sends $U$
and $V$ to $(id, \sigma)$ and $(\sigma,id)$, respectively, where
$\sigma\in S_d$ is a cyclic permutation of $d$, which is a
particular example of a permutation without fixed points.

\begin{theorem}
\label{symmetriesANDorthogonality} Let $({\mathcal E},{\mathcal
F})$ be a MUB pair and let $\mathbf b$ be a  normalized vector
that is unbiased to both of them. Let $U_0=\mathbf 1, U_1,\ldots ,
U_k$ be unitary operators implementing symmetries
$\alpha_0:=\alpha_{U_0}=id, \alpha_1:=\alpha_{U_1},\ldots,
\alpha_k:=\alpha_{U_k}$ of $({\mathcal E},{\mathcal F})$. Assume
further that for every $j\neq l$ the image of
$\alpha_j^{-1}\circ\alpha_l$ under the natural injection into
$S_n\times S_n$ is of the form $(\sigma, id)$ or $(id,\sigma)$
where $\sigma\in S_n$ is a permutation with no fixed points. Then
$$
(U_0\mathbf b = \mathbf b, U_1\mathbf b, \ldots, U_k \mathbf b)
$$
is an orthonormal family of vectors that are unbiased to both
${\mathcal E}$ and ${\mathcal F}$.
\end{theorem}

\begin{proof}
Suppose that the image of $\alpha_j^{-1}\circ\alpha_l$ under the
natural injection into $S_d\times S_d$ is of the form $(\sigma,
id)$. Then each vector of $\mathcal F$ must be an eigenvector for
$U_j^*U_l$: there exist some $\lambda_1,\ldots,\lambda_d \in
\mathbb C$ such that $U_l^*U_j \mathbf{f}_k = \lambda_k
\mathbf{f}_k$. Thus
\begin{eqnarray*}
\nonumber \langle  U_l \mathbf b, U_j \mathbf b \rangle &=&
\langle \mathbf b, U_l^*U_j \mathbf b \rangle = \sum_k \langle
\mathbf b, \mathbf f_k \rangle \langle\mathbf f_k,U_l^*U_j \mathbf
b \rangle \\ \nonumber &=& \sum_k \langle \mathbf b,\mathbf f_k
\rangle \langle U_j^*U_l  f_k, \mathbf b \rangle = \sum_k
\lambda_k
|\langle \mathbf b, \mathbf f_k \rangle|^2\\
& =&\frac{1}{d}{\rm Tr}\,(U_j^*U_l).
\end{eqnarray*}
For this last identity we have used the fact that $\mathbf b$ is
unbiased to $\mathcal F$, thus $|\langle \mathbf b, \mathbf f_k
\rangle|^2 = 1/d$, and the fact that the sum of the eigenvalues of
a diagonalizable operator is its trace. However, by assumption
$U_j^*U_l$ takes the vector $\mathbf e_k$ into a multiple of
$\mathbf e_{\sigma(k)}$; say to $\mu_k\mathbf{e}_{\sigma(k)}$.
Therefore
$$
{\rm Tr}\,(U_j^*U_l) = \sum_k \langle U_j^*U_l \mathbf e_k,
\mathbf e_k\rangle = \sum_k \langle \mu_k e_{\sigma(k)} , \mathbf
\mathbf e_k \rangle = 0
$$
as $\mathbf e_k$ is always orthogonal to $\mathbf e_{\sigma(k)}$
(since $\sigma$ has no fixed points).
\end{proof}

\subsection{Application to the case of ${\mathcal F}_{(a,b)}$}

We shall now apply the general statements made so far to the case
$({\mathcal E},{\mathcal F}(a,b))$ where $\mathcal{E}$ is the
standard basis of $\C^6$. As it was explained, we have numerical
evidence that up to multiple, the number of common normalized
unbiased vectors is always $48$. At $(a,b)=(0,0)$, this is a known
fact.

\begin{theorem}
There exists a neighbourhood $K$ of $(0,0)$ such that when counted
up to multiples, $({\mathcal E},{\mathcal F}_{(a,b)})$ has exactly
$48$ common normalized unbiased vectors for all $(a,b)\in K$.
\end{theorem}

\begin{proof}
For simplicity, Theorem \ref{UBcontinuity} and Corollary
\ref{UBnumberunchanged} were stated for a one-parameter smooth
pair of MUB. However, the proofs only rely on the Implicit
Function Theorem, so they easily extend to any number of
parameters.

But for $(\mathcal E,\mathcal F(0,0))$ {\it i.e.} for the standard
and the (usual) Fourier basis, all 48 vectors (counted up to
multiples) that are unbiased to them are explicitly known
\cite{BaBj}. It is then easy (but cumbersome) to check that the
conditions of Corollary \ref{UBnumberunchanged} hold for each of
them.
\end{proof}

We have seen that there is a theoretical reason (at least in a
neighbourhood of the origin) behind the numerical facts that
indicate that the number of common unbiased vectors (counted up to
multiples) to $(\mathcal{E},\mathcal{F}(a,b))$ is always $48$.
Unfortunately, we have so far been unable to find a theoretical
ground for the fact that these vectors can always be grouped into
$8$ orthonormal bases. However, we may now give a partial result
by applying what we have established about symmetries. To do so,
first we shall need to investigate in particular the symmetries of
the pair $({\mathcal E},{\mathcal F}_{(a,b)})$.

Consider the unitaries $U$ and $V$ defined by equation
(\ref{defUV}). For a generic value of the parameters $a,b$ they do
not implement symmetries. However, $U^2$ and $V^3$ implement
symmetries of $({\mathcal E},{\mathcal F}_{(a,b)})$ for all
$(a,b)\in \mathbb R^2$. Indeed, it is easy to check, that
regardless of the value of $a$ and $b$, we still have the
relations
\begin{eqnarray*}
\nonumber U^2\mathbf e_k = e^{i\frac{2\pi}{6}2(k-1)} \mathbf
e_k,\;\;\;\; &\textrm{and}& \;\;\;\; U^2\mathbf f_k = \mathbf
f_{k+2},\\
V^3\mathbf e_k = \mathbf e_{k-3} \;\;\;\; &\textrm{and}& \;\;\;\;
V^3\mathbf f_k = e^{i\frac{2\pi}{6}3(k-1)} \mathbf f_k,\;\;\;\;
\end{eqnarray*}
where now $\mathbf f_1,\ldots, \mathbf f_6$ are the vectors of
$\mathcal F_{(a,b)}$. Thus by applying Theorem
\ref{symmetriesANDorthogonality} we can draw the following
conclusion.

\begin{corollary}
Let $a,b$ be two fixed real numbers. Suppose $\mathbf b$ is a
common unbiased vector to the standard basis $\mathcal{E}$ and to
the basis $\mathcal{F}_{(a,b)}$. Consider the unitaries $U$ and
$V$ defined by equation (\ref{defUV}). Then all of the vectors in
the table below
\begin{center}
  \begin{tabular}{ | c | c | c | }
    \hline
&\ & \\[-8pt]
    $\mathbf b$ & $U^2 \mathbf b$ & $U^4\mathbf b$ \\[3pt]
\hline
&\ & \\[-8pt]
    $V^3 \mathbf b$ & $U^2 V^3 \mathbf b$ & $U^4 V^3 \mathbf b$ \\[3pt]
    \hline
  \end{tabular}
\end{center}
are unbiased to both bases. Moreover, each row and each column
consists of pairwise orthogonal vectors.
\end{corollary}

Unfortunately, this does not show that every common normalized
unbiased vector can be extended to an ONB consisting of common
unbiased vectors, only. However, in particular it shows that every
common normalized unbiased vector can be extended to an
orthonormal {\it triplet} of unbiased vectors.

\newpage

\appendix

\section{The calculations leading to Theorem \ref{thmmubtriplet}}

In this section we provide the detailed calculations leading to
Theorem \ref{thmmubtriplet}.

Recall the form of the Fourier matrices $F(0,b)$ from equation
\eqref{fab}, with $x=1$, $y=e^{2i\pi b}$. We will first look for
vectors ${\bf
u}=\frac{1}{\sqrt6}(1,\overline{c}_1,\overline{c}_2,\overline{c}_3,\overline{c}_4,\overline{c}_5)$.
 that obey the following further constraints\footnote{This particular
form was suggested by numerical evidence.}: $c_1=c_3c_4$ or
equivalently $c_1\overline{c_4}=c_3$.

We will also write $c_5=\eta\zeta$ and $c_2=\overline{\eta}\zeta$
with $|\eta|=|\zeta|=1$. Then Lemma \ref{lem:ph1117} implies that
\begin{eqnarray}
\mbox{Re}(2c_3+\eta^2)&=&0\notag\\
(1+\mbox{Re}\,c_3)\overline{c_4}+c_4(1+c_3)\overline{\zeta}\mbox{Re}\,\eta
+\overline{(1+c_3)}\zeta\mbox{Re}\,\eta&=&0\label{eq:ph2-a}\\
(1-\mbox{Re}\,c_3)\overline{c_4}+i\overline{y}c_4(1-c_3)\overline{\zeta}\mbox{Im}\,\eta
-iy\overline{(1-c_3)}\zeta\mbox{Im}\,\eta&=&0.\label{eq:ph3-a}
\end{eqnarray}
We will write $c_4=c_6^2$ and multiply \eqref{eq:ph2-a},
\eqref{eq:ph3-a} by $\overline{c_6}$ to obtain
\begin{eqnarray}
(1+\mbox{Re}\,c_3)\overline{c_6}^3+2\mbox{Re}\,\bigl(\overline{(1+c_3)c_6}\zeta\bigr)\mbox{Re}\,\eta
&=&0\label{eq:ph2}\\
(1-\mbox{Re}\,c_3)\overline{c_6}^3+2\mbox{Im}\,\bigl(\overline{(1-c_3)c_6}\zeta
y\bigr)\mbox{Im}\,\eta &=&0.\label{eq:ph3}
\end{eqnarray}

Notice that either \eqref{eq:ph2} or \eqref{eq:ph3} imply that
$c_6^3$ is real {\it i.e.} $c_6^3=\pm1$. We will restrict our
attention to $c_6^3=1$, that is $c_6=1,\omega,\omega^2$, thus
$c_4=c_6^2=\overline{c_6}=1,\omega^2$ or $\omega$.

Further, writing $\nu=\overline{c_6}\zeta$, $z=-iy$ and using
elementary computations, we obtain
\begin{eqnarray}
2\mbox{Re}\,c_3+\mbox{Re}\,\eta^2&=&0\label{eq:pha}\\
(1+\mbox{Re}\,c_3)\bigl(1+2\mbox{Re}\,\nu\mbox{Re}\,\eta\bigr)
+2\mbox{Im}\,c_3\mbox{Im}\,\nu\mbox{Re}\,\eta
&=&0\label{eq:phb}\\
(1-\mbox{Re}\,c_3)\bigl(1+2\mbox{Im}\,(\nu
y)\,\mbox{Im}\,\eta\bigr) +2\mbox{Im}\,c_3\mbox{Re}\,(\nu
y)\,\mbox{Im}\,\eta&=&0. \label{eq:phc}
\end{eqnarray}

\medskip

Assume now we have a solution $(c_3,\nu,\eta)$ of the system
\eqref{eq:pha}-\eqref{eq:phb}-\eqref{eq:phc}. Then
$(1,c_3c_4,\nu\overline{\eta c_4},c_3,c_4,\nu\eta\overline{c_4})$
with $c_4=1,\omega$ or $\omega^2$ are solutions
of\eqref{eq:1c}-\eqref{eq:1d}-\eqref{eq:1e} (with $x=1$). In other
words, the conjugates (!) of the following vectors
$$
w_1=(1,c_3,\nu\bar\eta,c_3,1,\nu\eta),\quad
w_2=(1,c_3\omega,\nu\bar\eta\omega^2,c_3,\omega,\nu\eta\omega^2)
$$
and
$$
w_3=(1,c_3\omega^2,\nu\bar\eta\omega,c_3,\omega^2,\nu\eta\omega)
$$
are unbiased to both the standard and the $F(0,b)$ basis. It is
also easy to see that these three vectors are orthogonal to each
other.

However, we need three more vectors. This is achieved via a {\it
miracle} that was indicated by numerical evidence. More precisely,
assume that $(c_3,\eta,\nu)$ is a solution of the system and that
$(\tilde c_3,\tilde\eta,\tilde\nu)=(-c_3,i\eta,\tilde\nu)$ is
another solution of the system. We then have 6 vectors that are
unbiased to both the standard basis and to $F(0,b)$. Moreover, the
three vectors $\overline{w_1},\overline{w_2},\overline{w_3}$
stemming from the first solution are orthogonal, and so are the
ones stemming from the second solution,
$\overline{w_4},\overline{w_5},\overline{w_6}$, namely
$$
w_4=(1,-c_3,-i\tilde\nu\bar\eta,-c_3,1,i\tilde\nu\eta),\quad
w_5=(1,-c_3\omega,-i\tilde\nu\bar\eta\omega^2,-c_3,\omega,i\tilde\nu\eta\omega^2)
$$
and
$$
w_6=(1,-c_3\omega^2,-i\tilde\nu\bar\eta\omega,-c_3,\omega^2,i\tilde\nu\eta\omega).
$$
Finally, it is easy to check that each of
$$\overline{w_1},\overline{w_2},\overline{w_3}$$ is orthogonal to
each of $$\overline{w_4},\overline{w_5},\overline{w_6}$$ so that
$(\overline{w_1},\ldots, \overline{w_6})$ is an orthogonal basis
unbiased to both the standard and the $F(0,b)$ basis.

\medskip

It thus remains to exhibit two such families of solutions. To be
more precise, we will write $\eta=e^{it}$ and show that, for a
certain range of $t$, we may chose $y=e^{i\beta(t)}$ in such a way
that the system \eqref{eq:pha}-\eqref{eq:phb}-\eqref{eq:phc} has a
solution $\bigl(c_3(t),e^{it},\nu(t)\bigr)$, and such that there
is a second solution $\bigl(-c_3(t),ie^{it},\tilde\nu(t)\bigr)$.

Now, if $\eta=e^{it}$, then $\mbox{Re}\,c_3(t)=-\frac{\cos 2t}{2}$
and, as $|c_3|=1$, there are only two possibilities,
$c_3(t)=-\frac{\cos 2t}{2}\pm
i\left(1-\frac{\cos^22t}{4}\right)^{1/2}$. For sake of simplicity,
we will take the $+$ sign:
\begin{equation}
\label{def:c3} c_3(t)=-\frac{\cos 2t}{2}+
i\left(1-\frac{\cos^22t}{4}\right)^{1/2}.
\end{equation}
Let us first determine $\nu=e^{i\ffi(t)}$. To reduce the length
and complexity of formulas, we will drop the dependence on $t$ in
them and simply write $c_3,\beta,\ffi$.

But $\nu$ satisfies \eqref{eq:phb}-\eqref{eq:phc} which now read
\begin{eqnarray*}
-(2-\cos2t)\cos t\cos\ffi-\sqrt{4-\cos^22t}\cos t\sin\ffi
&=&1-\frac{\cos2t}{2}\\[6pt]
\ \bigl((2+\cos2t)\sin\beta+\sqrt{4-\cos^22t}\cos\beta\bigr)\sin t\cos\ffi&&\\
+\bigl((2+\cos2t)\cos\beta-\sqrt{4-\cos^22t}\sin\beta\bigr)\sin
t\sin\ffi &=&-1-\frac{\cos2t}{2}
\end{eqnarray*}
Let us write these equations in a simpler form by introducing the
following parameter:
\begin{equation}
\label{def:psichi} \psi=\arccos\frac{\sqrt{2+\cos 2t}}{2}
\end{equation}
so that $\cos\psi=\frac{\sqrt{2+\cos 2t}}{2}$ and
$\sin\psi=\frac{\sqrt{2-\cos 2t}}{2}$. A simple computation then
shows that we want to solve
\begin{eqnarray}
-\sin\psi\cos t\cos\ffi-\cos\psi\cos t\sin\ffi
&=&\dst\frac{\sin\psi}{2}\label{eq:simple1}\\
\sin\bigl(\beta+\psi\bigr)\sin
t\cos\ffi+\cos\bigl(\beta+\psi\bigr)\sin t\sin\ffi
&=&\dst-\frac{\cos\psi}{2}\label{eq:simple2}
\end{eqnarray}

\begin{remark}\rm
This system may not have solutions. For instance, it is easy to
see that $\cos\psi$ and $\sin\psi$ do not vanish, but the left
hand side of \eqref{eq:simple1} ---resp. \eqref{eq:simple2}---
vanishes when $t=\pi/2$ --- resp. $t=0$. So for $t$ near $0$ or
$t$ near $\pi/2$, we do not expect to find a solution this way.

\end{remark}

The solution is now easy to obtain:
\begin{equation}
\label{defcffi} \cos\ffi=\frac{-\cos^2\psi\cos
t+\cos(\beta+\psi)\sin\psi\sin t} {\sin\beta\sin 2t}
\end{equation}
and
\begin{equation}
\label{defsffi} \sin\ffi=\frac{\sin\psi\cos\psi\cos
t-\sin(\beta+\psi)\sin\psi\sin t} {\sin\beta\sin 2t}.
\end{equation}

It still has to be shown that this is a legitimate solution, that
is, to check whether \eqref{defcffi}-\eqref{defsffi} define the
cosine and sine of an angle $\ffi$. For this, it is sufficient to
check that $\cos\ffi$, $\sin\ffi$ defined by these formulas
satisfy $\cos^2\ffi+\sin^2\ffi=1$. This easily reduces to
\begin{eqnarray}
&&\cos^2\psi\cos^2t+\sin^2\psi\sin^2t-2\cos t\sin t\cos\psi\sin\psi\cos\beta\notag\\
&&=\sin^2\beta\sin^2 2t. \label{eq:c2s2}
\end{eqnarray}
Note that $\cos\psi\sin\psi=\dst\frac{\sqrt{4-\cos^22t}}{4}$, and
\begin{eqnarray*}
&&\cos^2\psi\cos^2t+\sin^2\psi\sin^2t-\sin^22t\\
&=&\frac{2+\cos 2t}{4}\cos^2t+\frac{2-\cos
2t}{4}\sin^2t-\sin^22t=-\frac{1}{2}+\frac{5}{4}\cos^2 2t.
\end{eqnarray*}
We thus have to check that
$$
-\frac{1}{2}+\frac{5}{4}\cos^2 2t-\frac{\sqrt{4-\cos^22t}}{4}
u+u^2=0
$$
where $u=\cos\beta\sin 2t$. One solution of this equation is
\begin{eqnarray}
\cos\beta&=&\frac{\sqrt{4-\cos^22t}+3\sqrt{4-9\cos^22t}}{8\sin 2t}\notag\\
&=& \frac{\sqrt{3+\sin^22t}+3\sqrt{9\sin^22t-5}}{8\sin
2t}\label{eq:beta}
\end{eqnarray}
and we omit the possible other root here. It is clear that,
whenever $\frac{\sqrt{5}}{3}\leq \left|\sin 2t\right|\leq 1$
holds, we obtain a legitimate real number for $\beta$.

It remains to find $\tilde\nu=e^{i\tilde\ffi}$ such that
$$
\bigl(\tilde
c_3(t),\tilde\eta(t),\tilde\nu(t)\bigr)=\bigl(-c_3(t),ie^{it},\tilde\nu(t)\bigr)
$$
is also a solution of \eqref{eq:pha}, \eqref{eq:phb},
\eqref{eq:phc}. Recall, that the value of
$y=\textbf{e}^{\textbf{i}\beta}$ has just been determined.

Note that
$$
2\mbox{Re}\,\tilde c_3+\mbox{Re}\,\tilde{\eta}^2=-\bigl(
2\mbox{Re}\,c_3+\mbox{Re}\,\eta^2\bigr)=0
$$
so that \eqref{eq:pha} is satisfied. The other two equations read
$$
-(2+\cos 2t)\sin t\cos\tilde{\varphi}+\sqrt{4-\cos^2 2t}\sin
t\sin\tilde{\varphi}=-\frac{2+\cos 2t}{2}
$$
and
$$
((2-\cos 2t)\sin\beta-\sqrt{4-\cos^22t}\cos\beta)\cos
t\cos\tilde{\varphi}+
$$
$$
+((2-\cos 2t)\cos\beta+\sqrt{4-\cos^22t}\sin\beta)\cos
t\sin\tilde{\varphi}=-\frac{2-\cos 2t}{2}
$$
where the dependence on $t$ in $\beta$ and $\tilde\ffi$ has been
dropped. From this, we deduce that
\begin{eqnarray}
\cos{\tilde\varphi} &=&\frac{-\sin t\sin^2\psi+\cos\psi\cos
t\sin(\beta+\psi)}{\sin\beta\sin 2t}
\label{cffi2}\\
\sin{\tilde\varphi} &=&\frac{\cos \psi\cos(\beta+\psi)\cos
t-\cos\psi\sin t\sin\psi}{\sin\beta\sin 2t}. \label{sffi2}
\end{eqnarray}

It is left to see that $\tilde\varphi$ is a legitimate real
number, that is summing the squares of the two numbers defined in
\eqref{cffi2}-\eqref{sffi2} yields $1$. It is easy to check that
this holds if and only if \eqref{eq:c2s2} holds, hence there are
no further restrictions on $t$.

In summary, we have proved Theorem \ref{thmmubtriplet}

\section{A construction by G. Zauner that leads to another one-parameter family}
This section is inspired by G. Zauner's PhD thesis \cite{Za}. As
this thesis is only available in German, we take this occasion to
present his construction to a wider audience and to compare his
construction to our family given in Theorem \ref{thmmubtriplet}.
We emphasize that the all credit for the results of this section
goes to G. Zauner.

\medskip

Let us recall that a \emph{circulant} matrix $A$ is a matrix of
the form
$$
A=\begin{pmatrix}
a_0    &a_1     &\cdots &\cdots   &a_{m-1}\\
a_{m-1}&a_0     &a_1    &\cdots   &\\[3pt]
\vdots &\ddots&\ddots&\ddots&\vdots\\
\vdots&        &\ddots&a_0&a_1\\
a_1    &\cdots  &\cdots  &a_{m-1} &a_0
\end{pmatrix}.
$$
It is easy to check that, for each $k=0,\ldots,m-1$, the vector
$$
\mathbf{f}_k=(1,\omega^k,\omega^{2k},\ldots,\omega^{(m-1)k}),
\qquad \omega=e^{2i\pi/m}
$$
is an eigenvector of $A$. We may thus write $A=\mathcal{F}^*_m\bar
A\mathcal{F}_m$ where $\mathcal{F}_m=[m^{-1/2}\omega^{jk}]_{0\leq
j,k\leq m-1}$ is the $m\times m$ Fourier matrix, and $\bar A$ is a
diagonal matrix.

Now, let $A_{0,0},A_{0,1},A_{1,0},A_{1,1}$ be four $m\times m$
circulant matrices and write $A_{i,j}=\mathcal{F}^*_m\bar
A_{i,j}\mathcal{F}_m$ where
$$
\bar A_{i,j}=\mbox{diag}\,(\alpha_{i,j}(0),\ldots\alpha_{i,j}(m-1)):=\begin{pmatrix}\alpha_{i,j}(0)&&0\\
&\ddots&\\ 0&&\alpha_{i,j}(m-1)\end{pmatrix}
$$
is diagonal. Let $T$ be the $2m\times 2m$
matrix given by $T=\begin{pmatrix}A_{0,0}&A_{0,1}\\
A_{1,0}&A_{1,1}\end{pmatrix}$. Then
$$
T=\begin{pmatrix}\mathcal{F}^*_m&0\\
0&\mathcal{F}^*_m\end{pmatrix}
\begin{pmatrix}\bar A_{0,0}&\bar A_{0,1}\\ \bar A_{1,0}&\bar A_{1,1}\end{pmatrix}
\begin{pmatrix}\mathcal{F}_m&0\\ 0&\mathcal{F}_m\end{pmatrix}
$$
so that $T$ is unitary if and only if $\begin{pmatrix}\bar
A_{0,0}&\bar A_{0,1}\\ \bar A_{1,0}&\bar A_{1,1}\end{pmatrix}$ is
unitary. But, this matrix is unitary if and only if the $m$
matrices $S_k=\begin{pmatrix}\alpha_{0,0}(k)&\alpha_{0,1}(k)\\
\alpha_{1,0}(k)&\alpha_{1,1}(k)\end{pmatrix}$ are unitary ($k=0,
\dots, m-1$). Finally, one may easily check that a $2\times 2$
matrix is unitary if and only if it can be written in the form
$$
S(\beta_0,\beta_1,\beta_2,\beta_3)=\frac{1}{2}\begin{pmatrix}
e^{i\beta_0}+e^{i\beta_1}&e^{i\beta_3}(e^{i\beta_0}-e^{i\beta_1})\\
e^{i\beta_2}(e^{i\beta_0}-e^{i\beta_1})&e^{i\beta_2}e^{i\beta_3}(e^{i\beta_0}+e^{i\beta_1})
\end{pmatrix}.
$$
For all $0\le k \le m-1$ we may thus write
$S_k=S\bigl(\beta_0(k),\beta_1(k),\beta_2(k),\beta_3(k)\bigr)$ and
define
$U_\ell=\mbox{diag}\,(e^{i\beta_\ell(0)},\ldots,e^{i\beta_\ell(m-1)})$
for $\ell=0,1,2 $ and $3$. Then define
\begin{equation}
E_1= \frac{1}{\sqrt{2}}
\begin{pmatrix}\mathcal{F}_m&U_2^*\mathcal{F}_m\\\mathcal{F}_m&-U_2^*\mathcal{F}_m\end{pmatrix}
\label{zaun1}
\end{equation}
and
\begin{equation}
E_2=\frac{1}{\sqrt{2}}\begin{pmatrix}U_0\mathcal{F}_m&U_0U_3\mathcal{F}_m\\U_1\mathcal{F}_m&-U_1U_3\mathcal{F}_m\end{pmatrix},
\label{zaun2}
\end{equation}
and a straightforward computation gives $T=E_1^*E_2$. Finally,
note that $E_1$ and $E_2$ are Hadamard matrices so that, if $T$
itself is a Hadamard matrix, then the standard matrix, the columns
of $E_1$ and the columns of $E_2$ are three mutually unbiased
bases in $\C^{2m}$.

As an example for $m=3$, Zauner \cite{Za} considers the following
matrix:
$$
T(x)=\frac{1}{\sqrt{6}}\begin{pmatrix}1&-e^{-ix}&e^{ix}&-1&ie^{-ix}&ie^{ix}\\
e^{ix}&1&-e^{-ix}&ie^{ix}&-1&ie^{-ix}\\
-e^{-ix}&e^{ix}&1&ie^{-ix}&ie^{ix}&-1\\
1&ie^{-ix}&ie^{ix}&1&e^{-ix}&-e^{ix}\\
ie^{ix}&1&ie^{-ix}&-e^{ix}&1&e^{-ix}\\
ie^{-ix}&ie^{ix}&1&e^{-ix}&-e^{ix}&1
\end{pmatrix}.
$$
Then $T(x)$ is a one-parameter family of complex Hadamard matrices of the form $T(x)=\begin{pmatrix}A_{0,0}(x)&A_{0,1}(x)\\
A_{1,0}(x)&A_{1,1}(x)\end{pmatrix}$. Therefore, the construction
above yields a one-parameter family of MUB-triplets $(Id, E_1(x),
E_2(x)).$

Finally we note that Zauner's family $\left( Id, E_1(x), E_2(x)
\right)$ is not equivalent to our family presented in Theorem
\ref{thmmubtriplet}. This can be seen in the following way. After
dephasing the rows and columns the transition matrix
$T(x)=E_1^*(x)E_2(x)$ is easily seen to be a member of the
Dita-family $D_6(x)$ (cf. \cite{karol} for the Dita-family of
complex Hadamard matrices of order 6). However, in our
construction in Theorem \ref{thmmubtriplet}, generically none of
the appearing matrices $F(0,b(t)), C(t)$ and $F(0,b(t))^*C(t)$ are
members of the Dita-family. This is true, because $F(0,b(t)),
C(t)$ are members of the generalized Fourier family $F(a,b)$,
while the transition matrix $F(0,b(t))^*C(t)$ generically has a
much larger Haagerup-invariant set than the Dita-matrices
$D_6(x)$, therefore they cannot be equivalent.



\newpage


\begin{thebibliography}{11}
\bibitem{AE}
\textsc{Y. Aharonov \& B.-G. Englert},
\newblock{\em The mean king's problem: Spin 1.}
Z. Naturforsch. {\bf 56a}, (2001) 16.

\bibitem{BBRV}
\textsc{S. Bandyopadhyay, P.\,O. Boykin, V. Roychowdhury \& F. Vatan}
\newblock{\em A New Proof for the Existence of Mutually Unbiased Bases.}
Algorithmica {\bf 34} (2002), 512–-528.

\bibitem{BN}
\textsc{K. Beauchamp \& R. Nicoara},
\newblock{\em Orthogonal maximal Abelian $\ast$-subalgebras of the $6\times6$ matrices}.
Linear Algebra Appl. {\bf 428} (2008), 1833--1853.

\bibitem{BPT}
\textsc{H. Bechmann-Pasquinucci \& W. Tittel},
\newblock{\em Quantum cryptography using larger alphabets.}
Phys. Rev. A, {\bf 61} (2000), no. 6, 062308, 6 pp.

\bibitem{BBELTZ}
\textsc{I. Bengtsson, W. Bruzda, \AA. Ericsson, J.-A. Larsson, W.
Tadej \& K. \.Zyczkowski},
\newblock{\em Mutually unbiased bases and Hadamard matrices of order six.}
J. Math. Phys. {\bf 48} (2007), no. 5, 052106, 21 pp.

\bibitem{BB}
\textsc{C. H. Bennett \& G. Brassard},
\newblock{\em Quantum cryptography: Public key distribution and coin tossing.}
In Proceedings of the IEEE Intl. Conf. Computers, Systems, and
Signal Processing, pages 175--179. IEEE, 1984.

\bibitem{BaBj}
\textsc{G. Bj\"orck \& B. Saffari},
\newblock{\em New classes of finite unimodular sequences with unimodular Fourier transforms. Circulant Hadamard matrices with complex entries.}
C. R. Acad. Sci. Paris, Serie 1 {\bf 320} (1995), 319--324.

\bibitem{BP}
\textsc{A. Bonami \& J.-B. Poly}
\newblock{\em The discrete Pauli problem}.
In preparation.

\bibitem{config}
\textsc{S. Brierley \& S. Weigert},
\newblock{\em Maximal sets of mutually unbiased quantum states in dimension six.}
arXiv:0808.1614 (quant-ph).

\bibitem{ujbrit}
\textsc{S. Brierley \& S. Weigert}, \newblock{\em Constructing
Mutually Unbiased Bases in Dimension Six.} arXiv:0901.4051 (2009)


\bibitem{numerical}
\textsc{P. Butterley \& W. Hall}
\newblock{\em Numerical evidence for the maximum number of mutually unbiased bases
in dimension six}.
Physics Letters A {\bf 369} (2007) 5–-8.

\bibitem{Com0}
\textsc{M. Combescure}
\newblock{\em The mutually unbiased bases revisited.}
 Adventures in mathematical physics,  29--43, Contemp. Math., {\bf 447}, Amer. Math. Soc., Providence, RI, 2007.

\bibitem{Com1}
\textsc{M. Combescure}
\newblock{\em Circulant matrices, Gauss sums and mutually unbiased bases I. The prime number case}.
Available at {\tt Arxiv:0710.5642v1.}

\bibitem{Com2}
\textsc{M. Combescure}
\newblock{\em Circulant matrices, Gauss sums and mutually unbiased bases II. The prime power case}.
Available at {\tt Arxiv:0710.5643v1.}

\bibitem{Cor}
\textsc{J.\,V. Corbett}
\newblock{\em The Pauli problem, state reconstruction
and quantum-real numbers}
Reports On Mathematical Physics {\bf 57} (2006) 53--68.

\bibitem{DGS}
\textsc{P. Delsarte, J. M. Goethals \& J.J. Seidel},
\newblock{\em  Bounds for systems of lines, and Jacobi polynomials.}
Philips Res. Repts. (1975), 91--105.

\bibitem{grassl}
\textsc{M. Grassl},
\newblock{\em On SIC-POVMs and MUBs in Dimension 6}.
in: Proc. ERATO Conference on Quantum Information
Science (EQUIS 2004), J. Gruska (ed.)

\bibitem{haagerup}
\textsc{U. Haagerup},
\newblock{\em Ortogonal maximal Abelian $\ast$-subalgebras of $n\times n$ matrices and cyclic $n$-roots}.
Operator Algebras and Quantum Field Theory (Rome), Cambridge, MA
International Press, (1996), 296--322.

\bibitem{Ho}
\textsc{S. G. Hoggar},
\newblock{\em t-designs in projective spaces.}
Europ. J. Combin. {\bf 3} (1982), 233–-254.

\bibitem{Iv}
\textsc{I. D. Ivanovic},
\newblock{\em Geometrical description of quantal state determination.}
J. Phys. A {\bf 14} (1981), 3241.

\bibitem{Ja}
\textsc{Ph. Jaming},
\newblock{\em Phase retrieval techniques for radar ambiguity problems.}
J. Fourier Anal. Appl. {\bf 5} (1999), 309--329.

\bibitem{arxiv}
\textsc{Ph. Jaming, M. Matolcsi, P. M\'ora, F. Sz\"oll\H{o}si, M.
Weiner},
\newblock{\em A generalized Pauli problem and an infinite family of MUB-triplets in dimension
6.} http://arxiv.org/abs/0902.0882

\bibitem{KL}
\textsc{G. A. Kabatiansky \& V. I. Levenshtein},
\newblock{\em Bounds for packings on a sphere and in space.}
Problems of Information Transmission {\bf 14} (1978), 1-–17.

\bibitem{KR}
\textsc{A. Klappenecker \& M. R\"otteler}
\newblock{\em Constructions of Mutually Unbiased Bases.}
Finite fields and applications,  137--144, Lecture Notes in Comput. Sci., {\bf 2948}, Springer, Berlin, 2004.

\bibitem{lam}
\textsc{C.W.H. Lam, L. H. Thiel \& S. Swiercz},
 \newblock{\em The non-existence of finite projective planes of order 10.}
 Can. J. Math., Vol: XLI, (1989) 1117-1123.

\bibitem{Msz} \textsc{M. Matolcsi, F. Sz\"oll\H osi},
\newblock{\em Towards a classification of 6x6 complex Hadamard matrices.} Open
Systems \& Information Dynamics, {\bf 15}, Issue:2, (June 2008),
93-108.


\bibitem{Re}
\textsc{J. M. Renes},
\newblock{\em Equiangular spherical codes in quantum cryptography.}
Quantum Inf. Comput. {\bf 5} (2005), 81--92.

\bibitem{Sc}
\textsc{J. Schwinger},
\newblock{\em Unitary Operator Bases}.
Proc Nat. Acad. Sci. U.S.A. {\bf 46}, (1960) 560.

\bibitem{Skinner} \textsc{A. J. Skinner, V. A. Newell, R. Sanchez},
\newblock{\em Unbiased bases (Hadamards) for 6-level systems: Four ways from
Fourier.} arXiv:0810.1761 (2008)

\bibitem{karol}
\textsc{W. Tadej \& K. \.Zyczkowski},
\newblock{\em A concise guide to complex Hadamard matrices}.
Open Syst. Inf. Dyn. {\bf 13} (2006), 133--177.



\bibitem{wer}
\textsc{R. F. Werner},
\newblock{\em All teleportation and dense coding schemes. Quantum information and computation.}
J. Phys. A, {\bf 34} (2001), 7081--7094.

\bibitem{WF}
\textsc{W. K. Wootters \& B. D. Fields},
\newblock{\em Optimal state-determination by mutually unbiased measurements.}
Ann. Physics {\bf 191} (1989), 363--381.

\bibitem{Za}
\textsc{G. Zauner},
\newblock{\em Quantendesigns – Grundz\"uge einer nichtkommutativen Designtheorie.}
PhD thesis, Universit\"at Wien, 1999. (available at
http://www.mat.univie.ac.at/$\sim$neum/ms/zauner.pdf)

\bibitem{web} Documentation of results: http://www.math.bme.hu/$\sim$matolcsi/angpubl.html

\end{thebibliography}
\end{document}